\documentclass[twoside,11pt]{article}
\usepackage{jair, theapa, rawfonts}

\usepackage{algorithm}
\usepackage{algpseudocode}
\usepackage{subcaption}
\usepackage{newfloat}
\usepackage{adjustbox}
\usepackage{listings}
\usepackage{xcolor}
\usepackage{tikz}
\usepackage{mathtools} 
\usepackage{amsfonts} %
\usepackage{cleveref}

\usetikzlibrary{arrows, automata, positioning}
\tikzset{auto, >=stealth}
\tikzset{every edge/.append style={shorten >= 1pt}}
\tikzset{
    main node/.style={circle,draw,minimum size=1cm,inner sep=0pt},
}

\DeclareCaptionStyle{ruled}{labelfont=normalfont,labelsep=colon,strut=off} 
\floatstyle{ruled}
\newfloat{listing}{tb}{lst}{}
\floatname{listing}{Listing}

\jairheading{-}{-}{-}{-}{-}
\ShortHeadings{GLM2FSA: Automaton-Based Representation from GLM}
{Yang, Gaglione, Neary, \& Topcu}
\firstpageno{1}

\usepackage{macros}
\usepackage{amsthm}
\usepackage{graphicx, array, blindtext}
\usepackage{dblfloatfix}
\theoremstyle{definition}
\newtheorem{definition}{Definition}

\begin{document}

\title{
Automaton-Based Representations of Task Knowledge \\from Generative Language Models
}

\author{\name Yunhao Yang \email yunhaoyang234@utexas.edu \\
        \addr University of Texas at Austin, USA\\
        \AND
       \name Jean-Rapha\"el Gaglione \email jr.gaglione@utexas.edu \\
       \addr University of Texas at Austin, USA\\
       \AND
       \name Cyrus Neary \email cneary@utexas.edu \\
       \addr University of Texas at Austin, USA\\
       \AND
       \name Ufuk Topcu \email utopcu@utexas.edu\\
       \addr The University of Texas at Austin, USA\\
       }

\maketitle

\begin{abstract}
Automaton-based representations of task knowledge play an important role in control and planning for sequential decision-making problems.
However, obtaining the high-level task knowledge required to build such automata is often difficult.
Meanwhile, large-scale \glspl{glm} can automatically generate relevant task knowledge. 
However, the textual outputs from \glspl{glm} cannot be formally verified or used for sequential decision-making.
We propose a novel algorithm named \GLMtoDFA{}, which constructs a \gls{aut} encoding high-level task knowledge from a brief natural-language description of the task goal.
\GLMtoDFA{} first sends queries to a \gls{glm} to extract task knowledge in textual form, and then it builds an \gls{aut} to represent this text-based knowledge.
The proposed algorithm thus fills the gap between natural-language task descriptions and automaton-based representations, and the constructed \glspl{aut} can be formally verified against user-defined specifications.
We accordingly propose a method to iteratively refine the queries to the \gls{glm} based on the outcomes, e.g., counter-examples, from verification.
We demonstrate \GLMtoDFA{}'s ability to build and refine automaton-based representations of everyday tasks (e.g., crossing the road), and also of tasks that require highly-specialized knowledge (e.g., executing secure multi-party computation).
\end{abstract}
\glsresetall


\section{Introduction}

%


Automaton-based representations of high-level task knowledge play a key role in planning and learning in sequential decision-making.
Such knowledge may include the requirements a designer wants to enforce on an agent, or a priori task information about the agent and the environment in which it operates.
Automaton-based representations are useful in many applications,
such as lexical analysis of compilers \cite{Brouwer1998MythsAF,ArnaizGonzlez2018SeshatA}, 
reinforcement learning \cite{Zhang2021LearningAM,Fang2020MultiagentRL,Valkanis2020ReinforcementLI,Xu2020JIRP},
and program verification \cite{Vardi1986AnAA}.


Despite their utility in a range of applications, capturing high-level task knowledge in automata is not straightforward.
Automaton learning algorithms infer such knowledge through queries to a human expert or an automated oracle \cite{learning-automata}.
In general, these algorithms may require an excessive number of queries to a human, and it is often unclear how an automated oracle can be constructed in the first place.
Even in cases in which an oracle exists, either the learning algorithm or the oracle requires prior information, such as the set of possible actions available to the agent and the set of environmental responses, i.e., symbols relevant for the automaton construction.
It is often unclear how to obtain this information.
Furthermore, the soundness of the inferred automaton depends on the choice of symbols.


We argue---and provide a proof of concept---that recent advances in so-called large-scale \glspl{glm} can help automatically distill high-level task knowledge into automaton-based representations.
Existing \glspl{glm}, such as the Generative Pre-trained Transformer (GPT) series \cite{brown2020GPT3}, are capable of generating realistic, human-like text in response to queries.
Such text often encodes rich world knowledge.
On the other hand, the outputs of \glspl{glm} are typically in a textual form that cannot be directly utilized for sequential decision-making or automaton learning.
Moreover, such textual outputs are not formally verifiable against user requirements, and so they cannot be used directly in safety-critical applications where correctness matters.

\begin{figure}[t]
    \centering
    \includegraphics[width=0.8\linewidth]{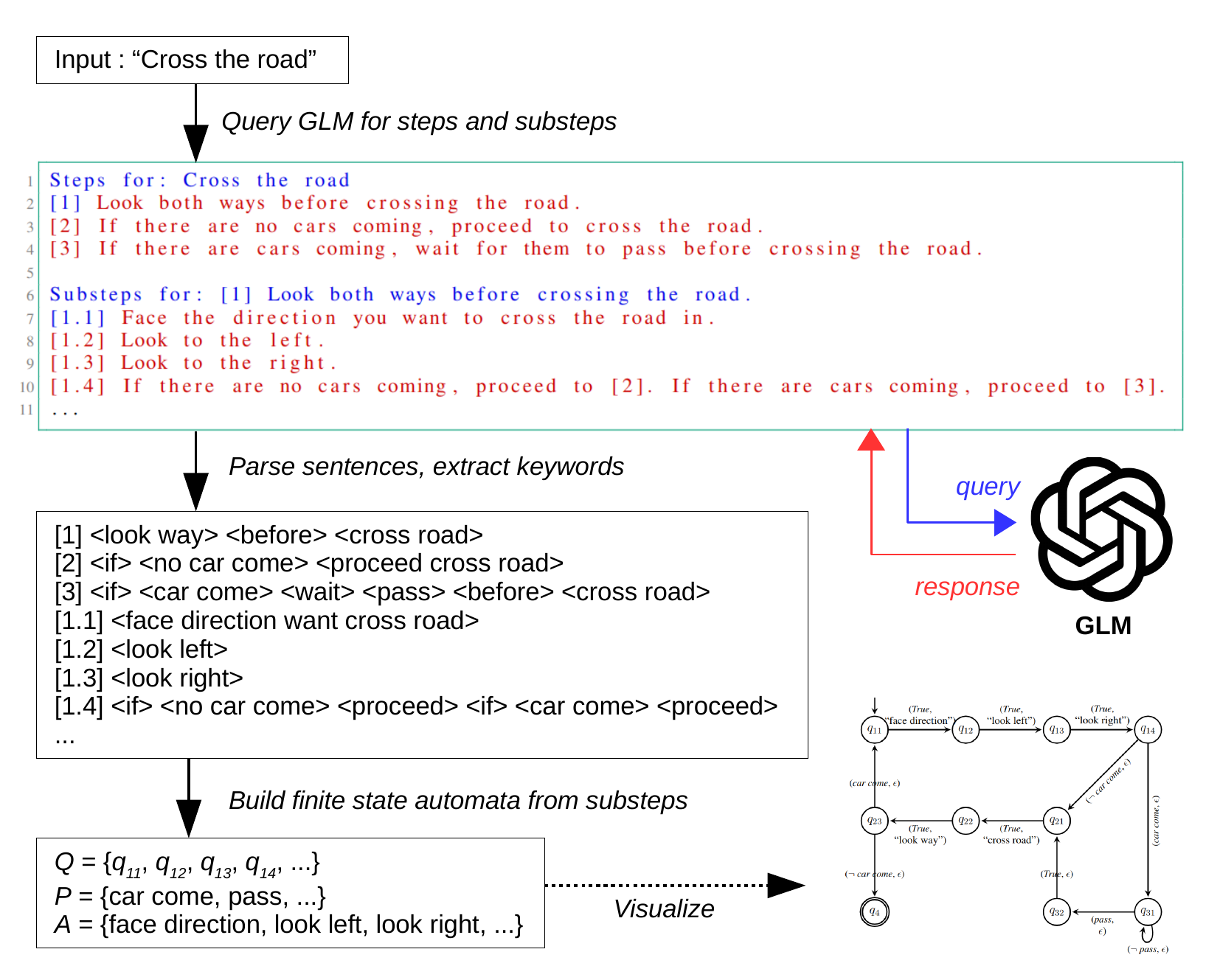}
    \caption{
    An illustration of the major steps in the \GLMtoDFA{} algorithm.
    A brief natural-language description is used to query the \gls{glm}, which outputs a series of steps in textual form.
    The algorithm then parses these textual outputs into verb phrases and connective keywords, which it uses to define the atomic propositions, output symbols, and transitions of the \glsfmtshort{aut} encoding the steps required to complete the task.
    A more detailed view of the output \glsfmtshort{aut} for this example is presented in Figure \ref{fig: substep-cross}.
    }
    \label{fig: exam}
\end{figure}

We develop an algorithm named \GLMtoDFA{} to fill the gap between the outputs from \glspl{glm} and automaton-based representations of high-level task knowledge. 
In particular, \GLMtoDFA{} produces controllers represented as \glspl{aut} from a brief natural-language sentence describing the task (e.g., ``cross the road").
It does so by first sending queries containing the task description to a \gls{glm} to obtain a list of text instructions organized in steps (and substeps).
Then, it parses these textual instructions to define the input and output symbols (i.e., environment propositions and actions) of the \gls{aut}.
Finally, it interprets each step to construct a corresponding automaton state and its outgoing transitions.
\GLMtoDFA{} thus constructs \glspl{aut} representing \textit{controllers} for sequential decision-making.
Figure \ref{fig: exam} illustrates the proposed \GLMtoDFA{} algorithm, which only takes a brief textual description of the task and outputs an \gls{aut} with automatically-defined states, symbols, and transitions.

The \gls{aut}-based controllers output by \GLMtoDFA{} are formally verifiable against user-defined task specifications. 
We accordingly propose a method to verify the controllers and to use the results of verification, e.g., counterexamples, as feedback in order to iteratively refine them through additional queries to the \gls{glm}. 
Such systematic verification allows the algorithm to identify and guard against potentially undesirable or nonsensical outputs from the language model, making it a necessary step towards the safe integration of \glspl{glm} into automated decision-making systems.

To the best of our knowledge, \GLMtoDFA{} is the first algorithm to construct automaton-based representations from textual knowledge extracted from large-scale \glspl{glm}.
It is also the first algorithm to provide an approach to formally verify the knowledge from \glspl{glm} in the context of sequential decision-making, and to use the results of the verification procedure to refine the extracted \glspl{aut}.

We demonstrate \GLMtoDFA{}'s capabilities through a variety of experimental case studies.
Experiments on both commonsense tasks (e.g., \textit{cross the road}) and tasks requiring highly-specialized knowledge (e.g., \textit{execute secure multi-party computation}) demonstrate the algorithm's ability to automatically distill domain-specific knowledge into control-oriented automaton-based representations.
It is able to do so even when only provided brief natural-language task descriptions, and when applied to highly specialized problem domains.
We additionally provide examples of the distilled controller failing the aforementioned verification step and triggering the refinement procedure.
In doing so, we demonstrate that the algorithm yields controllers that have verifiable properties even though their task descriptions are provided in the form of natural-language sentences.

\section{Related Work}

\paragraph{Extracting Task Knowledge from Language Models.}
Prior works have studied the extraction of task knowledge from language models \cite{xiong2019pretrained,davison2019commonsense,petroni2019language}.
However, due to a lack of rich world knowledge, the language models they use cannot generate action plans without providing detailed task descriptions.

Recently introduced large generative models for text---in particular the GPT series of \glspl{glm}---contain rich world knowledge and can generate instructions for a given task \cite{hendrycks2020measuring,evlp}.
Therefore, some works extract task-relevant knowledge by asking a \gls{glm} for the step-by-step instruction to solve a particular task of interest \cite{KnowledgeGraph,huang2022language}.

Meanwhile, a number of recent works have studied how recent advancements in the capabilities of \glspl{glm} can be used to extract task-relevant semantic knowledge and to generate plans for task completion in the context of robotics \cite{vemprala2023chatgpt,brohan2023can,shah2022robotic,huang2022language,huang2022inner}.
In contrast to existing works, we are the first to use \glspl{glm} to transform natural-language task descriptions into automaton-based representations that can be directly used for sequential decision-making and that can be formally verified against user-defined specifications.

\paragraph{Symbolic Knowledge Representations.}
Many works focus on constructing symbolic representations of task knowledge from natural language (text) descriptions. Several works extract information from text descriptions of given tasks and use that information to construct task-relevant knowledge graphs\cite{KnowledgeGraph,Rezaei2022UtilizingLM,He2022AcquiringAM}. 
Meanwhile, \citeA{Lu2022NeuroSymbolicCL} analyze the causality within the textual descriptions and create causal graphs.
\citeA{HumanRobotDialog} develop a semantic parser to map words to semantic forms, in order to help robots understand human dialog. 
Finally, \citeA{LogicExplanation} build logic-based representations of the solutions to planning tasks in order to explain why a particular plan is optimal.
In contrast with the existing works, we take advantage of the generative capabilities of \glspl{glm} to automatically generate automaton-based representations from brief (one-sentence) textual task descriptions.
Furthermore, the automaton-based representations we produce are directly applicable in algorithms for sequential decision-making and reinforcement learning \cite{icarte,Xu2020JIRP,Neider2021AdviceGuidedRL,neary2020reward,Fang2020MultiagentRL}.

\paragraph{Natural Language to Formal Language. }
Existing works introduce approaches to transform natural language to formal language specifications \cite{Vadera1994FromET,Baral2011UsingIL,Sadoun2013FromNL,Ghosh2014ARSENALAR}. 
\citeA{kate2005learning} induce transformation rules that map natural-language sentences into a formal query or command language.
\citeA{huang2022language} constructs a form of actionable knowledge that machines can recognize and operate on. 
However, existing works either cannot operate sequentially or cannot handle conditional transitions, e.g., multiple transitions from one state, which the work we proposed is capable of.

\section{Preliminaries}

\subsection{Finite State Automata}

A \glsreset{aut}\gls{aut} is a tuple
$\Aut =
\langle
    \AutSymbsIn,
    \AutSymbsOut,
    \AutStates,
    \Autstate[init],
    \AutTransFunc
\rangle$
where
$\AutSymbsIn$ is the input alphabet (the set of input symbols),
$\AutSymbsOut$ is the output alphabet (the set of output symbols),
$\Autstate[init] \in \AutStates$ is the initial state,
and $\AutTransFunc: \AutStates \times \AutSymbsIn \times \AutSymbsOut \times \AutStates \to \{0,1\}$
is the transition function, which indicates that a transition exists when it evaluates to $1$.
Note that \glspl{aut} transitions are non-deterministic in our definition: if an agent is in state $\Autstate_i \in \AutStates$ with an input symbol $\Autsymbin \in \AutSymbsIn$, the agent can choose the output symbol and next state among the set
$\AutTransFunc(\Autstate_i,\Autsymbin) \coloneqq \left\{
    (\Autsymbout,\Autstate_j) \in \AutSymbsOut\times\AutStates
\middle|
    \AutTransFunc(\Autstate_i,\Autsymbin,\Autsymbout,\Autstate_j) = 1
\right\}$.

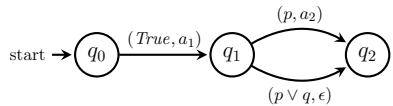
\begin{figure}[t]
    \centering
    \begin{tikzpicture}[
    scale=.6,
    node distance=2.2cm,
    thick,
    every node/.append style={transform shape},
]

\node[state,initial] (q0)
    at (3, 6)
    {\Large $\Autstate_{0}$};
\node[state] (q1)
    at (6, 6)
    {\Large $\Autstate_{1}$};
\node[state] (q2)
    at (9, 6)
    {\Large $\Autstate_{2}$};

\path[->,sloped]

(q0) 
edge[] node[]
    {$(\ltrue,\Autsymbout_1)$}
    (q1)

(q1) 
edge[bend left] node[]
    {$(\Autprop[1],\Autsymbout_2)$}
    (q2)
edge[bend right] node[swap]
    {$(\Autprop[1] \lor \Autprop[2],\noop)$}
    (q2)

;

\end{tikzpicture}
    \caption{
        An example \gls{aut}.
        Transitions are labeled with tuples $(\varphi,\Autsymbout)$, where $\varphi$ is a logic formula with atomic propositions in $\AutProps$, and $\Autsymbout$ is an output symbol in $\AutSymbsOut$.
        The transition from $\Autstate_0$ to $\Autstate_1$ happens unconditionally, with output symbol $\Autsymbout_1$.
        At state $\Autstate_1$, one can either choose output symbol $\Autsymbout_2$ or empty symbol ($\noop$), as long as the corresponding conditions ($\Autprop[1]$ and $\Autprop[1] \lor \Autprop[2]$, respectively) hold true.
    }
    \label{fig:aut-example}
\end{figure}

We use \glspl{aut} in the context of sequential decision-making,
where the input alphabet comprises all possible environment observations relevant to the current task.
We introduce a set of atomic proposition $\AutProps$ such that $\AutSymbsIn \coloneqq 2^{\AutProps}$, i.e., an input symbol $\Autsymbin \in \AutSymbsIn$ is the set of atomic propositions in $\AutProps$ that evaluate to \textit{True}.
We also introduce a set of atomic propositions $P_A$ for the output alphabets $\AutSymbsOut \coloneqq 2^{P_A}$.
Given a propositional logic formula based on these atomic propositions, e.g., $\varphi = \lnot\Autprop[1] \land \Autprop[2]$ where $\Autprop[1],\Autprop[2] \in \AutProps$,
the transition $(\Autstate_i,\varphi,\Autsymbout,\Autstate_j)$ exists
if for all $\Autsymbin \in \AutSymbsIn$ that satisfy $\varphi$ we have $\AutTransFunc(\Autstate_i,\Autsymbin,\Autsymbout,\Autstate_j) = 1$.
In words, if the current state is $\Autstate_i$ and condition $\varphi$ holds, one can choose to transition to the next state $\Autstate_j$ with output symbol $\Autsymbout$.
We also allow for a ``no operation'' or ``empty'' symbol $\noop \in \AutSymbsOut$.
Figure \ref{fig:aut-example} depicts an example \gls{aut}.

\subsection{Automaton-Based Representations of Controllers}
A controller is a system component responsible for making decisions and taking actions based on the system's state. A controller can be mathematically represented as a mapping from the system's current state to an action, which is executable in the task environment. 
In this work, we use \glspl{aut} output by the proposed algorithm to represent controllers. We refer to the input and output alphabets ($\AutSymbsIn$ and $\AutSymbsOut$) of the \gls{aut} as the \emph{condition set} and \emph{action set}, where $\Autsymbin \in \AutSymbsIn$ and $\Autsymbout \in \AutSymbsOut$ represent conditions and actions, respectively.

The controller's objective is to adjust the control input so that the system's state evolves in a way that accomplishes the designer's goals, while simultaneously satisfying certain performance criteria which might, for example, be specified using a formal language.




\subsection{Semantic Parsing}
Semantic parsing is a task in \emph{natural language processing} (NLP) that converts a natural language utterance to a logical form: A machine-understandable representation.

There are many approaches to semantic parsing and there are different types of logical forms.
We follow the approach that predicts part-of-speech (POS) tags for each token and that builds \emph{phrase structure} depending on \emph{phrase structure rules}, also known as a \emph{grammar}.
POS tags include noun (N), verb (V), adjective (AJD), adverb (ADV), etc. 
Phrase structures are a tree-structured logical form whose leaves are the POS tags of the given natural language utterance (i.e., sentence).
Phrase structure rules organize the POS tags into phrases like noun phrases (NP) and verb phrases (VP).
\begin{definition}
    A \emph{Noun Phrase (NP)} is a group of words headed by a noun. A \emph{Verb Phrase (VP)} is composed of a verb and its arguments. VP follows the following grammar:
    \begin{center}
        VP $\longleftarrow$ V VP

        VP $\longleftarrow$ V NP
    \end{center}
\end{definition}
The left-hand side of the grammar is composed of the components on the right-hand side. 
The grammar defines a verb phrase as being either composed of a verb and another verb phrase or composed of a verb and a noun phrase.

To standardize the words under the phrase structure, the parsing approach converts all the words to their original form, e.g., it removes singular or plural, past tense, etc. This operation eliminates cases where phrases with the same words in different tenses are categorized as being distinct.

\section{Methodology}

We propose an algorithm named \GLMtoDFA{}. The algorithm first uses a natural-language description of the task of interest to query the \gls{glm} and to obtain step-by-step task instructions in textual form.
It then automatically parses these text-based instructions to construct a controller, represented as an automaton, which can be used for sequential decision-making.

\begin{algorithm}[t]
  \caption{Query the \gls{glm} for Task Instructions}\label{alg:query}
  \begin{algorithmic}[1]
    \Procedure{\textbf{GLM2Step}}{String \textsc{task\_desc}, integer \textsc{depth}, List[String] keywords}
    \Comment{Obtain the instructions for a given task; depth is the number of layers of substeps.}
    \State \textbf{GLM}.bias = keywords
    \State \textsc{prompt} = \emph{``Steps for: "} + \textsc{task\_desc} + \emph{``$\backslash n$ [1]"}
    \State \textsc{answer} = \textbf{GLM}(\textsc{prompt}) 
    \State \textsc{step\_numbers} = [\emph{``[1]"}, \emph{``[2]"},...]
    \For{i in range(1, \textsc{depth})}
        \State \textsc{sub\_numbers} = []
        \State \textsc{answer} = []
        \For{number in \textsc{step\_numbers}}
            \State \textsc{sub\_prompt} = \emph{``Substeps for "}+number
            \State \textsc{answer}.append(\textbf{GLM}(\textsc{sub\_prompt})) 
            \State \textsc{sub\_numbers}.append(\emph{``[1.1]"},...)
        \EndFor
        
        \State \textsc{step\_numbers} = \textsc{sub\_numbers}
    \EndFor
    \State \textbf{return} \textsc{steps} = (\textsc{step\_numbers},  \textsc{answer})
    \EndProcedure
  \end{algorithmic}
\end{algorithm}

\subsection{Extracting Textual Knowledge}

The first step of the proposed approach is to distill task-relevant textual knowledge from the GLM by iteratively prompting it with structured natural-language queries. 
Given a task description of interest, the algorithm asks for steps to achieve the task.
We also provide a method to refine a step (or a substep) into its constituent substeps.
As an example, below we present the template of a conversation with \GPT{}, using the \emph{Davinci-002} model (prompts sent to \GPT{} are illustrated in blue and the corresponding completion of \GPT{} is shown in red):

\noindent\begin{minipage}{\linewidth}
\begin{lstlisting}[language=completion]
    <prompt>Steps for: <emph>task description</emph>
    [1]</prompt><completion> <emph>step description</emph>
    [<emph>step number</emph>] <emph>step description</emph>
    ...</completion>
    
    <prompt>Substeps for: [<emph>step number</emph>] <emph>step description</emph>
    [<emph>step number</emph>.1]</prompt><completion> <emph>substep description</emph>
    [<emph>substep number</emph>] <emph>substep description</emph>
    ...</completion>
\end{lstlisting}
\end{minipage}

Note that the prompts for substeps include the history of previously queried steps (prompt and completion).
Algorithm \ref{alg:query} depicts an iterative process that first queries the GLM for steps to accomplish the task description, and subsequently for substeps to accomplish these individual steps.
This iterative process allows for the automated decomposition of the task description into a structured hierarchy of steps and substeps, up to a pre-specified depth.
Alternatively, it could be used as a mechanism for the refinement of the GLM's output: Steps can be automatically broken into substeps if they are unclear or too difficult to accomplish without further guidance. 
This information for step refinement could come, for example, from a downstream verification algorithm or even from a human operator, both of which we will discuss further in Section \ref{sec:verification}.

\subsection{Building \Glspl{aut} from Textual Knowledge}

The next step in the proposed approach is parsing textual GLM outputs and constructing automata-based controllers. 
The step descriptions generated by the \gls{glm} are in textual form, which is not directly interpretable by planning or learning algorithms in sequential decision-making.
Algorithm \ref{algo:text2dfa}, which we refer to as \GLMtoDFA{}, transforms step descriptions from textual form to \glsentrylongpl{aut}, which resolves this interpretation problem.

The algorithm first applies semantic parsing to each step description to obtain the keywords and verb phrases.
The semantic parsing method classifies the part-of-speech tags \cite{Kucera1967ComputationalAO} of each word in a sentence and builds word dependencies based on natural language grammar.
The algorithm extracts the words whose part-of-speech tag is \emph{verb} and the dependencies of those words; each verb with its noun dependency is a verb phrase.

Next, the algorithm extracts pre-defined keywords from the sentence. 
These keywords belong to a pre-defined set of words that we use to define our grammar for automaton construction, such as \emph{if} and \emph{wait}. 
Examples of these keywords are highlighted as bold text in Table \ref{tab: grammar}.
In Algorithm \ref{algo:text2dfa}, we use \textbf{parse}\((\cdot)\) to denote the function that we implement to execute this keyword and verb phrase extraction process.
In our implementation of this step, we use the \textit{spaCy} library for semantic parsing \cite{spaCy}.

The algorithm interprets every parsed verb phrase as either a condition or an action.
We define these two categories of verb phrases more precisely as follows:
\begin{definition}
    \ACTVP{} is a verb phrase describing an action that the controller might take, and \CONVP{} is a verb phrase indicating the conditions for triggering the transitions in the FSA.
\end{definition}
For each verb phrase, the algorithm classifies it as \ACTVP{} by default, unless the grammar associated with the keywords specifies that it is a \CONVP{}, as illustrated in Table \ref{tab: grammar}.

If a verb phrase VP includes one or more of the words \textit{and}, \textit{or}, \textit{no}, or \textit{not}, then the algorithm applies further refinements to the VP as follows:

no/not VP$_1$ = $\neg$ VP$_1$,

VP$_1$ and VP$_2$ = VP$_1 \land$ VP$_2$,

VP$_1$ or VP$_2$ = VP$_1 \vee$ VP$_2$.

\begin{algorithm}[t]
  \caption{Natural Language to \Gls{aut}}\label{english2nfa}
  \begin{algorithmic}[1]
    \Procedure{\textbf{Step2\gls{aut}}}{function keyword\_handler, List[String] \textsc{steps}, List[String] keywords}
    \State $\AutStates$ = [a state for each step] + [absorbing state] 
    \State $\Autstate[init] = \AutStates[][0]$ 
    \State $\AutProps, \AutSymbsOut, \AutTransFunc = \{\}, \{\noop\}, \{\}$ 
    
    \For{state\_number in [$0 : |\AutStates| - 1$]}
        \State \textsc{s\_num} = step number of the current state
        \State \CONVP, \ACTVP, KEYS = \textbf{parse}(\textsc{steps[s\_num]}) 

        \If{any(keywords) in KEYS}
            \State keyword\_handler(\AutStates, \CONVP, \ACTVP, KEYS, keywords)
        \Else 
            \State create $\AutTransFunc(\Autstate_{\textsc{s\_num}}, \ltrue, \text{\ACTVP}, \Autstate_{\textsc{s\_num}+1})$
            \State $\AutSymbsIn, \AutSymbsOut \coloneqq \AutSymbsIn \cup VP^C, \AutSymbsOut \cup VP^A$
        \EndIf
    \EndFor
    
    \State \textbf{return} $\AutProps, \AutSymbsOut, \AutStates, \Autstate[init], \AutStates[final], \AutTransFunc$ 
    \EndProcedure
  \end{algorithmic}
  \label{algo:text2dfa}
\end{algorithm}

Next, the algorithm constructs an \gls{aut} from the steps and the verb phrases within these steps.
Recall that each step consists of a step number and a natural-language sentence.
It transforms the steps and verb phrases into the components of an \gls{aut}, including a finite set of states $\AutStates$, a finite set of atomic propositions $\AutProps$, a finite action set $\AutSymbsOut$, a transition function $\AutTransFunc$, and an initial state $\Autstate[init]$.

For each step, the algorithm adds a state $\Autstate_i$ representing the current step $i$ to $\AutStates$, adds \ACTVP{} to the set of output symbols $\AutSymbsOut$, and adds \CONVP{} to the set of atomic propositions $\AutProps$.
The algorithm defines the state corresponding to the first step as the initial state. It also adds an absorbing state after the state corresponding to the final step. The absorbing state only has one self-transition with input \textit{True} and output ``no operation".

\begin{table*}[t]
\centering
\begin{tabular}{m{0.12\textwidth} m{0.25\textwidth} m{0.28\textwidth} m{0.18\textwidth}}
\hline
Category & Grammar & Transition Rule & Example\\
\hline

\shortstack{Default \\ Transition} & \shortstack{\ACTVP} &
\vspace{0.2cm} \begin{tikzpicture}[thick,scale=.6, node distance=2.2cm, every node/.style={transform shape}]
	\node[state] (0) at (0, 0) {\Large $q_i$};
	\node[state] (1) at (4, 0) {\Large $q_{i+1}$};

	\draw[->, shorten >=1pt] (0) to[left] node[below, align=center] {$(\ltrue, \text{\ACTVP})$} (1);
\end{tikzpicture} & [dial number] \\
\hline

\shortstack{Direct \\ Transition} & \shortstack{\ACTVP $\:$ [j]} & \vspace{0.2cm}\begin{tikzpicture}[thick,scale=.6, node distance=2.2cm, every node/.style={transform shape}]
	\node[state] (0) at (0, 0) {\Large $q_i$};
	\node[state] (2) at (4, 0) {\Large $q_j$};

	\draw[->, shorten >=1pt] (0) to[left] node[below, align=center] {$(\ltrue,\noop)$} (2);
\end{tikzpicture} & [proceed] [1] \\
\hline

\shortstack{Conditional \\ Transition} & \shortstack{ \textbf{if} $\:$ \CONVP, $\:$ \ACTVP \\ \ACTVP $\:$ \textbf{if} $\:$ \CONVP} &
\vspace{0.2cm}\begin{tikzpicture}[thick,scale=.6, node distance=2.2cm, every node/.style={transform shape}]
	\node[state] (0) at (0, 1) {\Large $q_i$};
	\node[state] (1) at (3, 1) {\Large $q_{j}$};

	\draw[->, shorten >=1pt] (0) to[left] node[below, align=center, sloped] {$(\textit{\CONVP}, \text{\ACTVP})$} (1);
    \draw[->, shorten >=1pt] (0) to[loop left] node[align=center] {$(\lnot\textit{\CONVP}, \noop)$} ();
\end{tikzpicture} & [if] [no car], [cross] \\

\shortstack{Conditional \\ Transition \\ (if else)}& \shortstack{ \textbf{if} $\:$ \CONVP, $\: \text{\ACTVP}_{1}.$ \\ $\quad$ \textbf{if} $\: \neg$ \CONVP, $\: \text{\ACTVP}_{2}$ 
} &
\vspace{0.2cm}\begin{tikzpicture}[thick,scale=.6, node distance=2.2cm, every node/.style={transform shape}]
	\node[state] (0) at (0, 0) {\Large $q_i$};
	\node[state] (1) at (2, 0) {\Large $q_{j}$};
	\node[state] (2) at (-2, 0) {\Large $q_{k}$};

	\draw[->, shorten >=1pt] (0) to[left] node[below, align=center, sloped] {(\textit{\CONVP},\\ \ACTVP$_1$)} (1);
    \draw[->, shorten >=1pt] (0) to[left] node[below, align=center, sloped] {($\lnot\textit{\CONVP}$,\\ \ACTVP$_2$)} (2);
\end{tikzpicture} &
[if] [no car], [cross]. [if], [car] [stay]. \\
\hline

\shortstack{Self \\ Transition} & \shortstack{ \textbf{wait} $\:$  \CONVP $\:$ \ACTVP \\ \ACTVP $\:$  \textbf{after} $\:$ \CONVP } &
\vspace{0.2cm} \begin{tikzpicture}[thick,scale=.6, node distance=2.2cm, every node/.style={transform shape}]
	\node[state] (0) at (0, 0) {\Large $q_i$};
	\node[state] (1) at (2.5, 0) {\Large $q_{i+1}$};

	\draw[->, shorten >=1pt] (0) to[left] node[below, align=center] {(\textit{\CONVP}, \\ \ACTVP)} (1);
    \draw[->, shorten >=1pt] (0) to[loop left] node[align=center] {$(\lnot\textit{\CONVP}, \noop)$} ();
\end{tikzpicture} & [wait] [car pass] [cross]\\

& \shortstack{\ACTVP $\:$ \textbf{until} $\:$ \CONVP} &
 \begin{tikzpicture}[thick,scale=.6, node distance=2.2cm, every node/.style={transform shape}]
	\node[state] (0) at (0, 0) {\Large $q_i$};
	\node[state] (1) at (2.5, 0) {\Large $q_{i+1}$};

	\draw[->, shorten >=1pt] (0) to[left] node[below, align=center] {$(\textit{\CONVP}, \noop)$} (1);
    \draw[->, shorten >=1pt] (0) to[loop left] node[align=center] {$(\lnot\textit{\CONVP}, \text{\ACTVP})$} ();
\end{tikzpicture} & [stay] [until] [car pass]\\
\hline
\end{tabular}
\caption{Transition rules defined for keywords under a specific grammar.}
\label{tab: grammar}
\end{table*}

Next, the algorithm builds transitions between the states using the various transition rules that we define in detail below, and that we illustrate in Table \ref{tab: grammar}.

\paragraph{Default Transitions.}
We define default transitions as transitions from the current state to the next state with a condition $\ltrue$.
Each state $\Autstate_i$ only has one outgoing transition to its next state $\Autstate_{i+1}$, with the verb phrases from the $i^{th}$ step as the output symbols.
A default transition $\AutTransFunc(\Autstate_i,\ltrue,\text{VP}^A_i,\Autstate_{i+1})$ exists, unconditionally of the valuation of the atomic propositions in $\AutProps$ (hence the corresponding transition's condition is $\ltrue$).
A demonstration of the default transition is presented in the first row of Table \ref{tab: grammar}.


\paragraph{Direct State Transitions.}
Next, we define a direct state transition, which is a transition from the current state $\Autstate_i$ to a state other than the next state $\Autstate_{i+1}$.
The direct state transition happens when there is a verb phrase in the step description that contains the number corresponding to another step.
The algorithm builds a direct state transition from the current state to the state representing step $j$ with output symbol $\noop$ (no operation).

\paragraph{Conditional Transitions.} 
We also define a conditional transition, which is a transition that only happens when certain conditions are satisfied.
During automaton construction, the algorithm will build a conditional transition when a step description contains the keyword \emph{if}.
The conditions themselves are defined as one or more atomic propositions in $\AutProps$.

The algorithm builds two transitions from each sentence according to the patterns illustrated in Table \ref{tab: grammar}.
The first transition consists of a starting state $\Autstate_i$, a conjunction of atomic propositions \CONVP{}, a target state $\Autstate_j$, and a set of outputs \ACTVP{}.
If the \ACTVP{} does not lead to a direct state transition, then the transition will end at $\Autstate_{i+1}$. 
The second transition is a self-transition at $\Autstate_i$ with a condition $\lnot \text{\CONVP}$ and with an output symbol $\noop$ (no action).

\paragraph{Self-Transitions.}
Finally, we define a self-transition, whose starting and target states are identical.
The algorithm will construct self-transitions whenever a step description contains any of the keywords \emph{wait}, \emph{after}, or \emph{until}.
The self-transition will cause the automaton to stay in the current state until some logical condition is met, as specified by the keywords.
We accordingly use the keywords and surrounding verb phrases to define a conditional transition that breaks the self-transition loop and proceeds to the next state.

\section{Verification and Refinement}\label{sec:verification}

Automaton-based representations of knowledge are compatible with existing methods for formal verification against task specifications.
We now present an approach to apply such computational techniques to \GLMtoDFA{}'s outputs.
In doing so, we demonstrate that the algorithm yields controllers that have verifiable properties even though their task descriptions are provided in the form of natural-language sentences. 
We subsequently use the outputs of verification in order to automatically generate feedback which we use to revise and refine the prompts provided to the \gls{glm}.

\begin{figure}[t]
    \centering
    \includegraphics[width=0.8\linewidth]{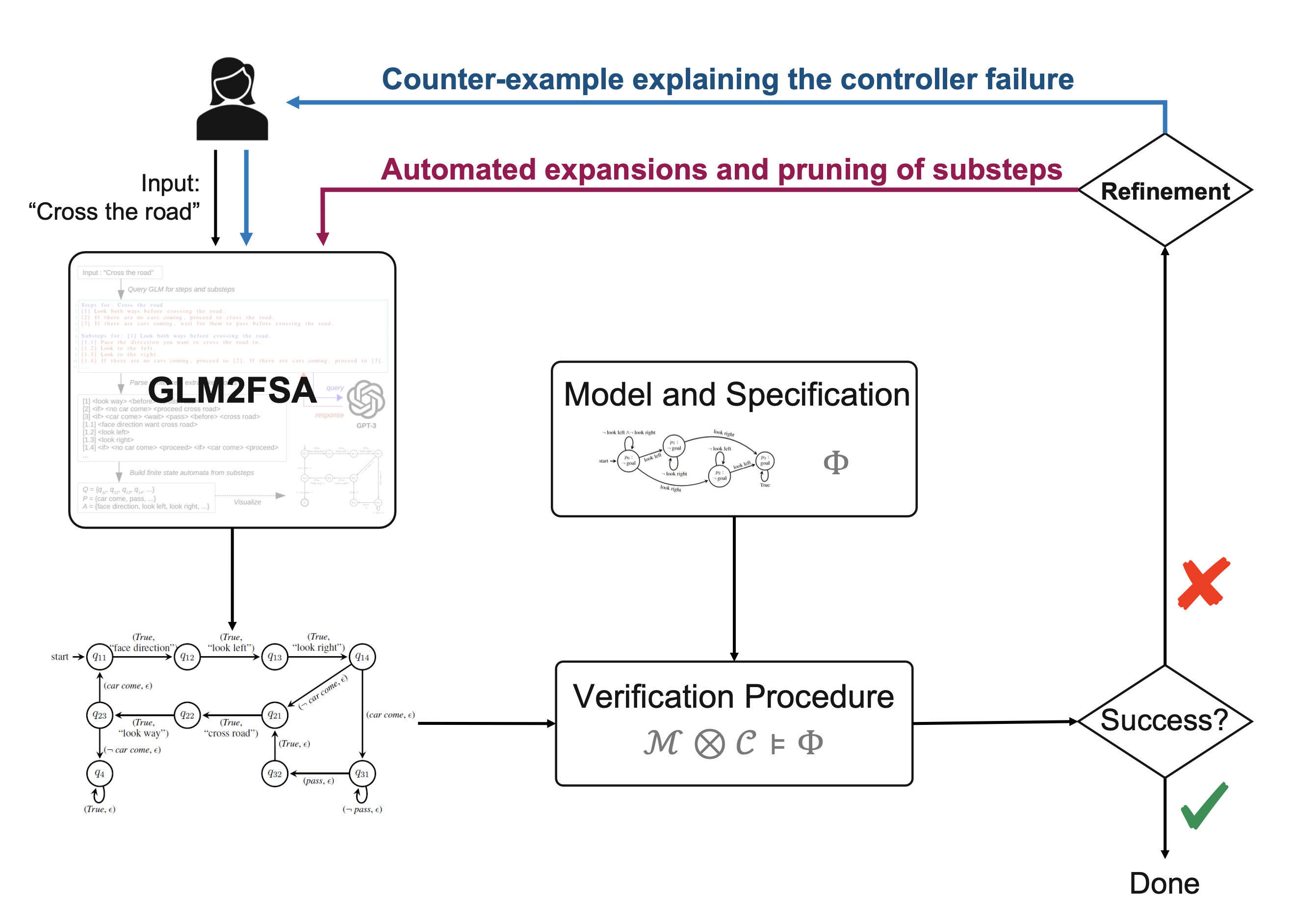}
    \caption{
        Illustration of the proposed procedure for automated verification and refinement. 
        We formally verify that the automaton output by \GLMtoDFA{} encodes behaviors that satisfy user-defined specifications.
        If the controller fails this verification step, we provide two methods to refine the \gls{glm} input prompts, and to update the generated controller accordingly.
    }
    \label{fig: refine}
\end{figure}

Figure \ref{fig: refine} provides an overview of the proposed approach for automated verification and automaton refinement.
Given an automaton-based controller $\Aut[controller]$ output by \GLMtoDFA{}, we use a \textit{model} $\Aut[model]$ to verify the behavior of $\Aut[controller]$ against some task specifications of interest.
If it fails to satisfy the task specification, we provide two separate methods to refine the input prompt to the \gls{glm} and to update \(\Aut[controller]\) accordingly. 
The first approach uses the counterexamples generated by the verification procedure as explanations of the controller's failure. These explanations are provided to the human user, who may subsequently ask the \gls{glm} to update the controller while accounting for the mistakes present in the counterexamples.
The second approach is fully automated and works by repeatedly expanding the FSA's steps into more detailed substeps and then by pruning the substeps that do not lead to a change in the result of the model checking problem.
We provide further details on the related model-checking problem, as well as on the two proposed approaches for refinement, below.

\subsection{Verifying the Automaton-Based Controllers Against Models and Task Specifications}
\label{sec:verification_subsection}
We begin with the problem of automatically checking that the behaviors of the controllers satisfy desired specifications when verified against a \emph{model}.
A \emph{model} is a representation of any a priori task-related knowledge provided by the user, e.g., an abstract model of how the task environment responds to the actions taken by the controller. 
The verification procedure checks if the controller is consistent with this existing knowledge or if it will satisfy logical-based specifications when implemented against the model.
Such systematic verification is necessary to identify and guard against undesirable or nonsensical outputs from the \gls{glm}.

Mathematically, we define the \emph{model} as a transition system
$\Aut[model] \coloneqq
\langle
    \AutSymbsIn[model],$
    $\AutSymbsOut[model],$
    $\AutStates[model],$
    $\Autstate[model-init],$
    $\AutTransFunc[model],$
    $\AutLabelFunc[model]$
$\rangle$.
Here, $\AutSymbsIn[model] \coloneqq \AutSymbsOut = 2^{P_A}$ is a set of input symbols for the model, where $P_A$ is a set of atomic propositions representing actions in the controller.
$\AutSymbsOut[model] \coloneqq 2^{ \{ goal \} \cup \AutProps}$ is a set of output symbols, where $\AutProps$ is the set propositions from $\Aut[controller]$ and \emph{goal} is a special additional proposition.
$\AutStates[model]$ is a finite set of states,
$\AutTransFunc[model]: \AutStates[model] \times \AutSymbsIn[model] \times \AutStates[model] \to \{0,1\}$ is a non-deterministic transition function, $\Autstate[model-init] \in \AutStates[model]$ is an initial state.
and $\AutLabelFunc[model]: \AutStates[model] \to \AutSymbsOut[model]$ is a labeling function.

We use \gls{ltl}~\cite{Pnueli77LTL} to define task specifications $\fLTL[spec]$ that the controller $\Aut[controller]$ should satisfy, given the model $\Aut[model]$.
\Gls{ltl} is a formal language that expresses system properties that evolve over time. 
It extends propositional logic by including temporal operators---$\leventually$ (``eventually'') and $\lalways$ (``always'')---which allow for reasoning about the system's future behavior.
We present the formal definition of \gls{ltl} in Appendix \ref{sec:appendix_background}, and we refer to Baier and Katoen \shortcite{baier2008principles} for further details.

We define specifications $\fLTL[spec]$ over atomic propositions in $\AutProps \cup P_A \cup \{\mathop{goal}\}$, and evaluate them over trajectories in the form $(2^{\AutProps \cup P_A \cup \{\mathop{goal}\}})^{*}$.
In the context of our verification problem, specifications $\fLTL[spec]$ represent desired outcomes that the controller should satisfy, given some assumptions on the properties of the system that the controller interacts with.

To verify that the controller $\Aut[controller]$ satisfies the specification $\fLTL[spec]$ given the model $\Aut[model]$, we solve the following automated verification problem,
\begin{equation}
    \Aut[model] \otimes \Aut[controller] \models \fLTL[spec],
    \label{eq:model-checking}
\end{equation}
where $\Aut[model] \otimes \Aut[controller]$ denotes the so-called \textit{product automaton} describing the interactions of the controller $\Aut[controller]$ with the model $\Aut[model]$.

Recall that the controllers \(\Aut[controller]\) output by GLM2FSA are defined as
$\Aut[controller] \coloneqq
\langle
    \AutSymbsIn,
    \AutSymbsOut,
    \AutStates,
    \Autstate[init],
    \AutTransFunc
\rangle$
with input alphabet $\AutSymbsIn := 2^{\AutProps}$, output alphabet $\AutSymbsOut := 2^{P_A}$,
and non-deterministic transition function $\AutTransFunc: \AutStates \times \AutSymbsIn \times \AutSymbsOut \times \AutStates \to \{0,1\}$.

We accordingly define the \emph{product automaton} to be a transition system
$\Aut[product] =
\Aut[model] \otimes \Aut[controller] \coloneqq
\langle
    \AutStates[product],$
    $\AutTransFunc[product],$
    $\Autstate[product-init],$    $\AutLabelFunc[product]
\rangle
$
with the following components:
\begin{align*}
\AutStates[product] &\coloneqq \AutStates[model] \times \AutStates
\\
\Autstate[product-init] &\coloneqq (\Autstate[model-init], \Autstate[init])
\\
\AutTransFunc[product]( (p,q) )
&\coloneqq
\left\{
    (p', q') \in \AutStates[product] 
    \middle|
    \AutTransFunc(q, \AutLabelFunc[model](p) \cap \AutSymbsIn, a, q') = 1
    \text{ and }
    \AutTransFunc[model](p, a, p') = 1, \text{for } a \in A
\right\}
\\
\AutLabelFunc[product]((p,q), (p',q')) 
&\coloneqq
\left\{
    \AutLabelFunc[model](p) \cup a
    \middle|
    a \in \AutSymbsOut
    \text{ and }
    \AutTransFunc(q, \AutLabelFunc[model](p)\cap \AutProps[model], a, q') = 1
    \text{ and }
    \AutTransFunc[model](p, a, p') = 1
\right\}.
\end{align*}

Here,
$\AutTransFunc[product] : \AutStates[product] \to 2^{\AutStates[product]}$
is a non-deterministic transition function,
and $\AutLabelFunc[product] : \AutStates[product] \times \AutStates[product] \to 2^{ \AutProps \cup P_A \cup \{goal\} }$ is a labeling function on the transitions of the product automaton.
The product automaton generates infinite  trajectories \((p_0,q_0), (p_1, q_1), \ldots\) by beginning in an initial state \(\Autstate[product-init]\) and following the nondeterministic transition function \(\AutTransFunc[product]\) thereafter. 
Labeled trajectories are then generated by applying the labeling function \(\AutLabelFunc[product]\) to these trajectories within the product automaton, i.e. \( \psi_0 \psi_1, \ldots \in ( 2^{ \AutProps \cup P_A \cup \{goal\} })^*\) where \(\psi_{i} \in \AutLabelFunc[product]((p_i, q_i), (p_{i+1},q_{i+1}))\).
When using the product automaton to solve the model-checking problem from Equation \eqref{eq:model-checking}, we check that all possible labeled trajectories generated by the product automaton belong to the language defined by the LTL specification.

We leave the details of the automated verification problem to Baier and Katoen \shortcite{baier2008principles}. However, we note that such problems can be efficiently solved using widely available and highly-optimized software libraries.
In this work, we use the NuSMV model checker~\cite{Cimatti2002NuSMV} for this purpose.
We present a detailed example of how we use the model checker to verify the behavior of an example controller against a user-specified model in the appendix.

The outcome of the automated verification problem is binary: The automaton-based controller $\Aut[controller]$ either satisfies the specification $\fLTL[spec]$ given the model or it does not.
However, as a byproduct of the verification procedure, if the controller fails to satisfy the specification, a counterexample is returned.
A counterexample returned by the model checker is a sequence of states from the product automaton that violates the specification. A counterexample is in a form of $(p_1, q_1), (p_2, q_2), ...$, where $p_i \in \AutStates[model]$ and $q_i \in \AutStates$. Intuitively, if an agent follows the sequence of states from the counterexample, it will eventually violate the specification.

Finally, we note that due to the stochastic nature of generative models, the \gls{glm} may often output different phrases to represent the same concept.
Furthermore, the verb phrases used to define the symbols of the externally provided model may use a different vocabulary than the one that was generated by the \gls{glm}.
This mismatch in verb phrases could lead to multiple distinct verb phrases being used to describe conditions and actions that we intuitively understand as being the same.
As a result, the automated verification procedure may yield unexpected failures because of its inability to recognize synonyms.

To remove these ambiguities, for every pair of verb phrases, we automatically query the \gls{glm} and ask it if the phrases have the same meaning.
More specifically, we use an input prompt in the following format for all pairs of verb phrases. 
\begin{lstlisting}[language=completion]
    <prompt> Do the two phrases "wait for the call to connect" and "wait for answer the call" lead to the same effect?</prompt>
    <completion> No. [specific reason......] </completion>
    <prompt> Do the two phrases "cross the road" and "walk across the road" lead to the same effect?</prompt>
    <completion> Yes. Both phrases lead to ... </completion>
\end{lstlisting}
If the \gls{glm} responds that two distinct verb phrases have the same meaning, we consolidate the phrases: We keep one of the phrases in our set of symbols and replace all instances of the other phrase.

\subsection{Refining the Automaton-Based Controllers}
\label{sec:refinement}

As illustrated in Figure \ref{fig: refine}, if the automaton output by \GLMtoDFA{} fails to satisfy the provided task specification, we provide two approaches to refine the controller $\Aut[controller]$.

\paragraph{Counterexample-Guided Controller Refinements.}
The first approach asks the user to manually modify the input prompt to the \gls{glm} and to use the resulting outputs to update the controller. 
In detail, if the verification steps fail, the model checker generates a counterexample. The counterexample is a sequence of states $(p_1, q_1), (p_2, q_2), ...$ from the product automaton, which can automatically be converted into a sequence of the product automaton's output labels \( \psi_0 \psi_1, \ldots \in ( 2^{ \AutProps \cup P_A \cup \{goal\} })^*\) where \(\psi_{i} \in \AutLabelFunc[product]((p_i, q_i), (p_{i+1},q_{i+1}))\).

The user can then interpret this counterexample information and use it to modify the \gls{glm}'s input prompt in a way that addresses the cause of the failure.
After creating this modified input prompt, the user can use \GLMtoDFA{} to construct the updated controller.
We provide detailed examples of this process, including illustrative counterexamples and the resulting user-specified refinements, in Section \ref{sec:experiments}.

\paragraph{Automated Refinement Through Substep Expansions.}
The second approach queries the \gls{glm} for substeps to automatically refine the controller $\Aut[controller]$.
We can again solve the verification problem for the newly-updated $\Aut[controller]$.
This approach leads to an iterative process that uses the information provided by the verification step to improve the distilled automaton.

During each refinement step, we apply Algorithm \ref{alg:query} to query the \gls{glm} for the next-layer steps (DEPTH = DEPTH + 1). 
This step expands each of the automaton's transitions into more detailed representations describing its necessary substeps.
We refer to the state that represents the beginning of the transition to be expanded as the \emph{parent state} and the states that represent that transition's substeps as the \emph{child states}.
We continue the loop of expanding the automaton's transitions and applying automated verification until all the specifications are satisfied or the maximum number of layers is reached.
This maximum number of layers is a user-defined constant.
If we reach the maximum number of layers and still cannot satisfy the specifications, we consider the task to be unrepresentable by an automaton.

Finally, we note that it is unlikely that all of the automata's steps need to be broken into substeps in order to satisfy the task specifications.
To prevent the above iterative refinement procedure from generating unnecessarily large automaton-based controllers, we thus propose an algorithm that uses the verification results to prune unnecessary states.
The pruning process is defined as follows: 1) Start from the deepest-layer steps and replace the first set of children states with their parent state, 2) check if the controller still satisfies all the specifications, 3) keep the controller as it is if the specifications are satisfied, otherwise add the children states back, and 4) continue steps 1 to 3 for all of the children states at each level of the hierarchy.

\section{Experimental Results}
\label{sec:experiments}

\subsection{Demonstrations of Controller Construction}
We construct several automaton-based controllers to demonstrate the capabilities of \GLMtoDFA{}. 
In particular, we provide examples to show that \GLMtoDFA{} can create \glspl{aut} that encode all of the knowledge required to describe the steps to complete a given task.
In all of the following experimental case studies, we use the \emph{text-davinci-003} model from the \GPT{} model family \cite{brown2020GPT3} as the \gls{glm} queried by \GLMtoDFA{}. 

\paragraph{Crossing the Road Example: A Sanity Check.}
To demonstrate the application of \GLMtoDFA{} to a relatively simple commonsense task, we begin with the task prompt ``cross the road," which is illustrated in Figure \ref{fig: exam}. 

First, we apply the algorithm to construct an \gls{aut} for \emph{first-layer} step descriptions. 
In this example, we query the \gls{glm} for two distinct scenarios: Crossing the road at a traffic light and crossing the road when there is no traffic light. 
The queries and the responses from the \gls{glm} are as follows:

\noindent\begin{minipage}{\linewidth}
\begin{lstlisting}[language=completion]
    <prompt>Steps for: Cross the road
    [1]</prompt><completion> Look both ways before crossing the road.
    [2] If there are no cars coming, proceed to cross the road.
    [3] If there are cars coming, wait for them to pass before crossing the road.</completion>
\end{lstlisting}
\begin{lstlisting}[language=completion]
    <prompt>Steps for: Cross the road at the traffic light
    [1]</prompt><completion> Locate the traffic light.
    [2] Wait for the traffic light to turn green.
    [3] Look both ways before crossing the road.
    [4] Cross the road if no cars are coming.</completion>
\end{lstlisting}
\end{minipage}

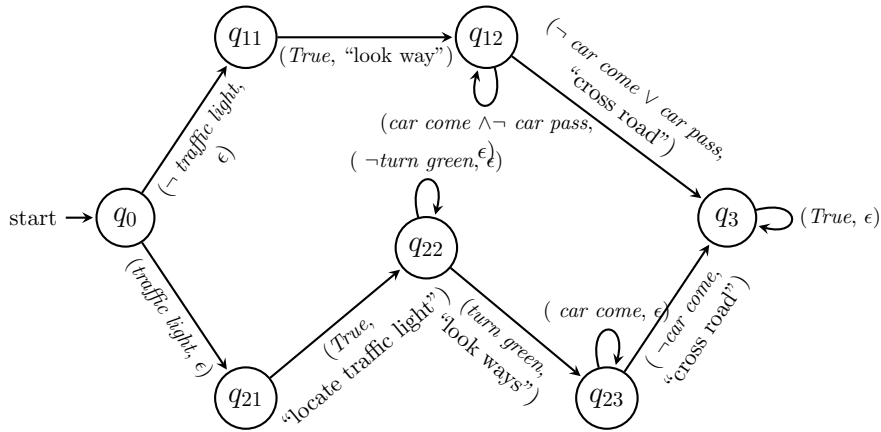
\begin{figure}[t]
    \centering
    \begin{tikzpicture}[thick,scale=.8, node distance=2.2cm, every node/.style={transform shape}]
	\node[state,initial] (0) at (0, 2) {\Large $q_0$};
	\node[state] (11) at (2, 5) {\Large $q_{11}$};
	\node[state] (12) at (6, 5) {\Large $q_{12}$};
    \node[state] (21) at (2, -1) {\Large $q_{21}$};
	\node[state] (22) at (5, 1.5) {\Large $q_{22}$};
    \node[state] (23) at (8, -1) {\Large $q_{23}$};
	\node[state] (4) at (10, 2) {\Large $q_3$};

	\draw[->, shorten >=1pt, sloped] (0) to[left] node[below, align=center] {\small ($\neg$ \textit{traffic light}, \\ $\noop$)} (11);
    \draw[->, shorten >=1pt, sloped] (0) to[left] node[below, align=center] {\small (\textit{traffic light}, $\noop$)} (21);
    
	\draw[->, shorten >=1pt] (11) to[left] node[below, align=center] {\small (\textit{True}, ``look way")} (12);
    \draw[->, shorten >=1pt] (12) to[left] node[above, align=center, sloped] {\small ($\neg$ \textit{car come $\vee$ car pass}, \\ ``cross road")} (4);
    
    \draw[->, shorten >=1pt] (21) to[left] node[below, align=center, sloped] {\small (\textit{True}, \\ ``locate traffic light")} (22);
    \draw[->, shorten >=1pt] (22) to[left] node[below, align=center, sloped] {\small (\textit{turn green}, \\ ``look ways")} (23);
    \draw[->, shorten >=1pt] (23) to[left] node[below, align=center, sloped] {\small ( $\neg$\textit{car come}, \\ ``cross road")} (4);
	
	\draw[->, shorten >=1pt] (12) to[loop below] node[align=center] {\small (\textit{car come $\land \neg$ car pass}, \\ $\noop$)} ();
    \draw[->, shorten >=1pt] (22) to[loop above] node[align=center] {\small ( $\neg$\textit{turn green}, $\noop$)} ();
    \draw[->, shorten >=1pt] (23) to[loop above] node[align=center] {\small ( \textit{car come}, $\noop$)} ();

    \draw[->, shorten >=1pt] (4) to[loop right] node[align=center] {\small (\textit{True}, $\noop$)} ();
\end{tikzpicture}
    \caption{
    An Automaton-based controller for the task ``cross the road'' that represents two ways of crossing the road depending on whether or not a traffic light is present.
    }
    \label{fig: cross}
\end{figure}

\GLMtoDFA{} constructs an \gls{aut} to represent these steps, illustrated in Figure \ref{fig: cross}. The constructed \gls{aut} successfully represents all the required knowledge---including the task-relevant actions and conditions---under the two different operating scenarios in which the traffic light either exists or does not exist.
The task steps are represented by states in the \gls{aut}.
The algorithm automatically generates the set of atomic propositions $\AutProps = \{ \textit{car come},$ $\textit{car pass},$ $\textit{turn green},$ $\textit{traffic light} \}$ and output symbols $A = \{$ ``look way," ``cross road," ``locate traffic light," $\noop \}$ from the extracted verb phrases.
Figure \ref{fig: cross} indicates that \GLMtoDFA{} is capable of building automaton-based representations that are unambiguous and able to represent task-relevant knowledge in first-layer step descriptions.

Next, we ask the algorithm to continue querying for the substeps of each step. 
We call these substep descriptions returned from \GPT{} \emph{second-layer} step descriptions.
To simplify the presentation, we only examine the second-layer descriptions for the scenario where no traffic light exists.

\begin{lstlisting}[language=completion]
    ...<!-- CompletionRequest(previous=root_step_request) -->
    <prompt>Substeps for: [1] Look both ways before crossing the road.
    [1.1]</prompt><completion> Face the direction you want to cross the road in.
    [1.2] Look to the left.
    [1.3] Look to the right.
    [1.4] If there are no cars coming, go to [2]. If there are cars coming, go to [3].</completion>
    
    <prompt>Substeps for: [2] If there are no cars coming, proceed to cross the road.
    [2.1]</prompt><completion> Walk across the road.
    [2.2] Once you have reached the other side, look both ways again to make sure no cars are coming.
    [2.3] If there are no cars coming, proceed to [4]. If there are cars coming, back to [1].</completion>
    
    <prompt>Substeps for: [3] If there are cars coming, wait for them to pass before crossing the road.
    [3.1]</prompt><completion> Wait for the cars to pass.
    [3.2] Once the cars have passed, back to [2].
\end{lstlisting}

\begin{figure}[t]
    \centering
    \begin{tikzpicture}[thick,scale=.8, node distance=2.2cm, every node/.style={transform shape}]
	\node[state,initial] (11) at (2, 4) {\Large $q_{11}$};
    \node[state] (12) at (6, 4) {\Large $q_{12}$};
    \node[state] (13) at (9, 4) {\Large $q_{13}$};
    \node[state] (14) at (12, 4) {\Large $q_{14}$};
 
	\node[state] (21) at (9, 2) {\Large $q_{21}$};
    \node[state] (22) at (6, 2) {\Large $q_{22}$};
    \node[state] (23) at (2, 2) {\Large $q_{23}$};

    \node[state] (31) at (12, 0) {\Large $q_{31}$};
    \node[state] (32) at (9, 0) {\Large $q_{32}$};
     
	\node[state] (4) at (2, 0) {\Large $q_4$};

	\draw[->, shorten >=1pt] (11) to[left] node[above, align=center] {\small (\textit{True}, \\ ``face direction")} (12);
	\draw[->, shorten >=1pt] (12) to[left] node[above, align=center, sloped] {\small (\textit{True}, \\ ``look left")} (13);
    \draw[->, shorten >=1pt] (13) to[left] node[above, align=center, sloped] {\small (\textit{True}, \\ ``look right")} (14);
    \draw[->, shorten >=1pt] (14) to[left] node[below, align=center, sloped] {\small (\textit{$\neg$ car come}, $\noop$)} (21);
    \draw[->, shorten >=1pt] (14) to[left] node[right, align=center] {\small (\textit{car come}, $\noop$)} (31);

    \draw[->, shorten >=1pt] (21) to[left] node[below, align=center, sloped] {\small (\textit{True}, \\ ``cross road")} (22);
    \draw[->, shorten >=1pt] (22) to[left] node[below, align=center, sloped] {\small (\textit{True}, \\ ``look way")} (23);
    \draw[->, shorten >=1pt] (23) to[left] node[right, align=center] {\small (\textit{car come}, $\noop$)} (11);
    \draw[->, shorten >=1pt] (23) to[left] node[right, yshift=-2mm, align=center] {\small (\textit{$\neg$ car come}, $\noop$)} (4);

	\draw[->, shorten >=1pt] (31) to[loop below] node[align=center] {\small (\textit{$\neg$ pass}, $\noop$)} ();
    \draw[->, shorten >=1pt] (31) to[left] node[below, align=center, sloped] {\small (\textit{pass}, $\noop$)} (32);
    \draw[->, shorten >=1pt] (32) to[left] node[right, align=center] {\small (\textit{True}, $\noop$)} (21);

    \draw[->, shorten >=1pt] (4) to[loop below] node[align=center] {\small (\textit{True}, $\noop$)} ();
\end{tikzpicture}
    \caption{An \gls{aut} that represents all the substeps for crossing the road when no traffic light is present.
    }
    \label{fig: substep-cross}
\end{figure}
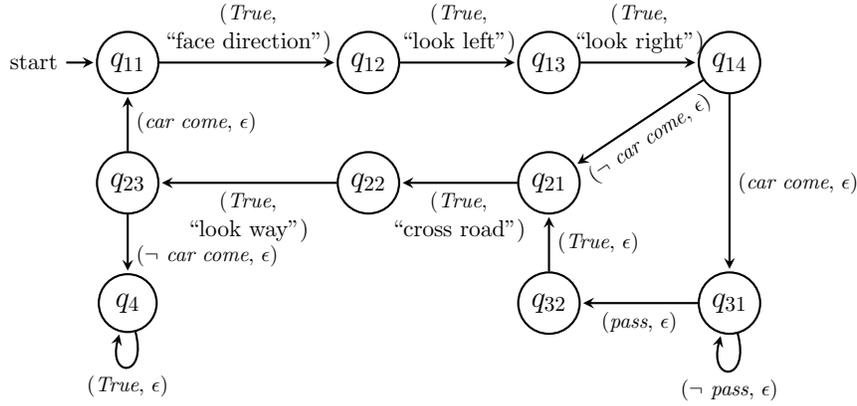

We present the constructed \gls{aut} for the second-layer descriptions in Figure \ref{fig: substep-cross}. 
The \gls{aut} consists of states representing each substep, a set of atomic propositions $\AutProps = \{ \textit{car come}, \textit{pass}\}$, output symbols $\AutSymbsOut = \{$``look way," ``cross road," ``face direction," ``look left," ``look right", $\noop \}$, an initial state $\Autstate_{11}$, and a final state $\Autstate_4$.
This example demonstrates the ability of \GLMtoDFA{} to construct \glspl{aut} that represent more detailed task descriptions.

We note that a small portion of the automaton in Figure \ref{fig: substep-cross} does not agree with what we might intuitively expect given the task description ``cross the road."
In particular, we expect that the action ``cross road" would lead us from state $\Autstate_{21}$ directly to the \textit{goal} state $\Autstate_{4}$: When we cross the road and no car is coming, the task is complete.
Instead, the automaton transitions from $\Autstate_{21}$ to $\Autstate_{22}$, where it again checks if a car is coming and potentially crosses the road a second time.

This observation highlights an important shortcoming of approaches that attempt to use large language models in autonomous systems directly: They may occasionally output instructions that are unexpected or nonsensical with respect to the intended task.
Methods to formally verify and refine the outputs of the language model are thus necessary for the development of algorithms that incorporate them into the design of autonomous systems.

\begin{figure}[t]
    \centering
    \begin{tikzpicture}[thick,scale=.7, node distance=2.2cm, every node/.style={transform shape}]
	\node[state,initial] (1) at (0, 4) {\Large $q_{1}$};
	\node[state] (2) at (4, 6) {\Large $q_{2}$};
    \node[state] (3) at (8, 6) {\Large $q_{3}$};
	\node[state] (4) at (12, 6) {\Large $q_{4}$};
	\node[state] (5) at (14, 5) {\Large $q_5$};

    \node[state] (11) at (0, 1) {\Large $q_{11}$};
    \node[state] (12) at (3, 1) {\Large $q_{12}$};
    \node[state] (13) at (4, 3) {\Large $q_{13}$};

    \node[state] (131) at (8, 3) {\Large $q_{131}$};
    \node[state] (132) at (12, 1) {\Large $q_{132}$};
    \node[state] (133) at (11, 4) {\Large $q_{133}$};

	\draw[->, shorten >=1pt, sloped, dashed] (1) to[bend left] node[above, align=center] {\small (\textit{True}, \\ ``research local clinic")} (2);

    \draw[->, shorten >=1pt, sloped] (2) to[left] node[above, align=center] {\small (\textit{True}, \\ ``read review")} (3);
    
	\draw[->, shorten >=1pt, color=blue] (1) to[left] node[above, align=center, sloped] {\small (\textit{True}, $\noop$)} (11);
    \draw[->, shorten >=1pt, color=blue] (11) to[bend right] node[below, align=center, sloped] {\small (\textit{True}, \\ ``online search...")} (12);
    \draw[->, shorten >=1pt, color=blue] (12) to[bend right] node[right, xshift=-6mm, align=center] {\small (\textit{True}, \\ ``gather recommendation...")} (13);
    \draw[->, shorten >=1pt, color=blue, dashed] (13) to[left] node[left, xshift=6mm, align=center] {\small (\textit{True}, \\ ``check insurance...")} (2);

    \draw[->, shorten >=1pt, sloped, color=purple] (13) to[left] node[above, align=center] {\small (\textit{True}, $\noop$)} (131);
    \draw[->, shorten >=1pt, sloped, color=purple] (131) to[left] node[above, align=center] {\small (\textit{True}, \\ ``get contact info")} (132);
    \draw[->, shorten >=1pt, sloped, color=purple] (132) to[bend right] node[above, align=center] {\small (\textit{True}, \\ ``call insurance...")} (133);
    \draw[->, shorten >=1pt, sloped, color=purple] (133) to[left] node[above, align=center] {\small (\textit{True}, ``request a list...")} (2);

    \draw[->, shorten >=1pt, sloped] (3) to[left] node[above, align=center] {\small (\textit{True}, \\ ``compare")} (4);
    \draw[->, shorten >=1pt, sloped] (4) to[bend left] node[above, align=center] {\small (\textit{True}, \\ ``schedule")} (5);

     \draw[->, shorten >=1pt] (5) to[loop below] node[align=center] {\small (\textit{True}, $\noop$)} ();
\end{tikzpicture}
    \caption{The \gls{aut}-based controller for finding a dentist and making an appointment. The {\color{blue} blue} transitions represent the second-layer substeps and the {\color{purple} purple} transitions represent the third-layer substeps. The dashed transitions are replaced by the transitions representing the substeps.}
    \label{fig: dental}
\end{figure}
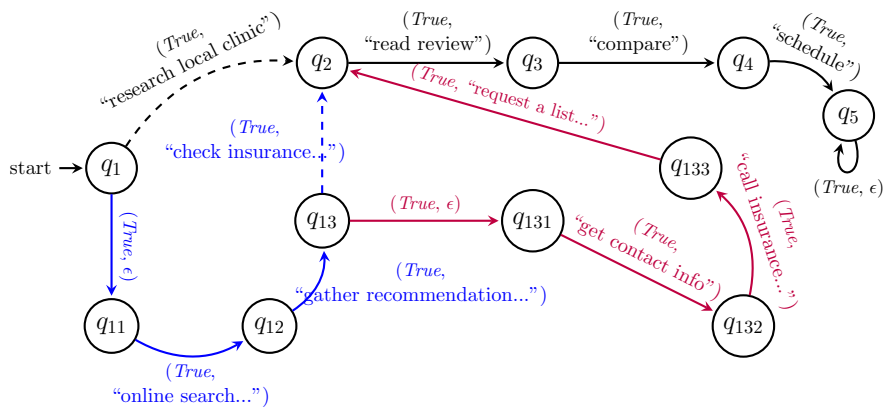

\paragraph{Booking a Dental Appointment Example: Hierarchical Expansions of Automaton Substeps.}
We have thus far shown the capability of \GLMtoDFA{} to generate first-layer and second-layer descriptions. 
In this example, we further explore the idea of expanding the (sub)steps and integrating the detailed substep descriptions from multiple layers of such expansions into the originally generated controller.
As an example, we use the task prompt: \textit{Find a dentist and make an appointment.}
Figure \ref{fig: dental} illustrates the resulting FSA and highlights the different levels of the hierarchical expansion in different colors.

In practice, we do not need the details for all the steps generated by \GPT{}. Some step descriptions are straightforward, while others may need further explanation. 
Therefore, we can expand the \gls{aut} to represent the first-layer, second-layer, and third-layer step descriptions only for the steps that need further explanations.

First, we apply the algorithm to query \GPT{} and obtain the step and substep descriptions:

\begin{lstlisting}[language=completion]
    <prompt>Steps for finding a dentist and making an appointment:
    [1]</prompt><completion> Research local dental clinics
    [2] Read patient reviews
    [3] Compare services and prices
    [4] Schedule an appointment</completion>

    <prompt>Substeps for step 1:
    [1.1]</prompt><completion> Online search for local dental clinics
    [1.2] Gather recommendations from acquaintances
    [1.3] Check insurance provider's in-network list</completion>

    <prompt>Substeps for step 1.3:
    [1.3.1]</prompt><completion> Get insurance provider's contact information
    [1.3.2] Call the insurance provider's customer service
    [1.3.3] Request a list of in-network dental clinics </completion>
\end{lstlisting}

Note that in this example, we only request substeps for step 1, whose action is ``research dental clinics," while the other steps are already straightforward. Then, we further expand the substeps for step 1.3 and obtain a list of thrid-layer substeps.

The algorithm builds three \glspl{aut} representing the three layers of substeps, respectively.
Then, we remove the outgoing transition from the state $\Autstate_1$ corresponding to the first step.
Instead, we build an outgoing transition from $\Autstate_1$ to the initial state of the \gls{aut} for second-layer substeps and replace the transition to the final state with the transition to state $\Autstate_2$.
Hence we create a partially extended \gls{aut}. Similarly, we connect the \gls{aut} to the third-layer substeps.
Figure \ref{fig: dental} illustrates the resulting extended \gls{aut}.

This example indicates that \GLMtoDFA{} can construct scalable \glspl{aut} representing detailed step descriptions. Every state in the constructed \gls{aut} can be extended to represent more details. We can continue expanding the substeps until reaching the detail level we expected following this pipeline. 

\paragraph{Secure Multi-Party Computation: Distilling Domain-Specific Task Knowledge.}

In addition to the daily-life tasks presented above, we now show that \GLMtoDFA{} with \GPT{} can also devise controllers for domain-specific tasks, which require highly specialized task knowledge. 
Such examples expand the application of \GLMtoDFA{} to fields where domain-specific human expertise would previously have been required.

In particular, we consider the task of ``secure multi-party computation." Secure multi-party computation (MPC) is a technique that allows multiple parties to jointly compute a function on their private inputs without revealing anything about their inputs to each other or to any other third party. 
MPC is a specialized problem in computer security, which may not be well-known by people outside this field. 

\begin{figure}[t]
    \centering
    \begin{tikzpicture}[thick,scale=.7, node distance=2.2cm, every node/.style={transform shape}]
	\node[state,initial] (1) at (0, 0) {\Large $q_{1}$};
	\node[state] (2) at (3, 0) {\Large $q_{2}$};
    \node[state] (21) at (3, -3) {\Large $q_{21}$};
    \node[state] (22) at (6, -3) {\Large $q_{22}$};
    
    \node[state] (3) at (6, 0) {\Large $q_{3}$};
    \node[state] (31) at (8, -3) {\Large $q_{31}$};
    \node[state] (32) at (11, -3) {\Large $q_{32}$};
    \node[state] (33) at (11, 0.5) {\Large $q_{33}$};
    \node[state] (34) at (11, 4) {\Large $q_{34}$};

	\node[state] (4) at (8, 2) {\Large $q_{4}$};
    \node[state] (5) at (6, 4) {\Large $q_{5}$};
	\node[state] (6) at (2, 4) {\Large $q_{6}$};
	\node[state] (7) at (0, 2) {\Large $q_7$};

	\draw[->, shorten >=1pt, sloped] (1) to[left] node[below, xshift=-1mm, align=center] {\small (\textit{True}, \\ ``define problem")} (2);
 
	\draw[->, shorten >=1pt, color=blue, dashed] (2) to[left] node[below, xshift=2mm, align=center, sloped] {\small (\textit{True}, $\noop$)} (21);
    \draw[->, shorten >=1pt, color=blue, dashed] (21) to[left] node[below, align=center, sloped] {\small (\textit{True}, \\ ``generate share")} (22);
    \draw[->, shorten >=1pt, color=blue, dashed] (22) to[left] node[below, align=center, sloped] {\small (\textit{True}, \\ ``store share")} (3);

    \draw[->, shorten >=1pt] (2) to[left] node[above, yshift=3mm, align=center, sloped] {\small (\textit{True}, \\ ``secret share")} (3);
    \draw[->, shorten >=1pt] (3) to[left] node[below, align=center, sloped] {\small (\textit{True}, \\ ``compute share")} (4);
 
    \draw[->, shorten >=1pt, color=blue, dashed] (3) to[left] node[below, align=center, sloped] {\small (\textit{True}, $\noop$)} (31);
    \draw[->, shorten >=1pt, color=blue, dashed] (31) to[left] node[below, align=center, sloped] {\small (\textit{True}, \\ ``encrypt share")} (32);
    \draw[->, shorten >=1pt, color=blue, dashed] (32) to[left] node[above, yshift=3mm, sloped, align=center] {\small (\textit{True}, \\ ``distribute share")} (33);
    \draw[->, shorten >=1pt, color=blue, dashed] (33) to[left] node[above, xshift=-2mm, align=center, yshift=4mm, sloped] {\small (\textit{True}, \\ ``compute ciphertext")} (34);
    \draw[->, shorten >=1pt, color=blue, dashed] (34) to[left] node[above, xshift=-1mm, yshift=3mm, align=center, sloped] {\small (\textit{True}, \\ ``broadcast result")} (4);
    
    \draw[->, shorten >=1pt] (4) to[left] node[below, xshift=-4mm, yshift=-3mm, align=center, sloped] {\small (\textit{True}, \\ ``reconstruct result")} (5);
    \draw[->, shorten >=1pt] (5) to[left] node[above, yshift=3mm, align=center, sloped] {\small (\textit{True}, \\ ``output verification")} (6);
    \draw[->, shorten >=1pt] (6) to[left] node[below, align=center, sloped] {\small (\textit{True}, \\ ``decrypt result")} (7);

     \draw[->, shorten >=1pt] (7) to[loop above] node[align=center] {\small (\textit{True}, $\noop$)} ();
\end{tikzpicture}
    \caption{Example \gls{aut} for secure multi-party computation. 
    The first-layer steps are represented by the states in black and the second-layer substeps are represented by the states whose transitions are in blue.
    }
    \label{fig: smpc}
\end{figure}
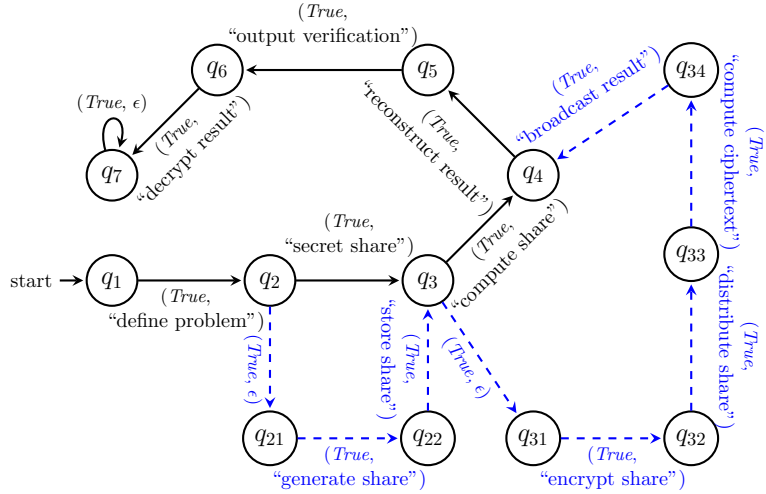

We query \GPT{} and get responses as the following:

\begin{lstlisting}[language=completion]
    <prompt>Steps for: Secure multi-party computation 
    [1]</prompt><completion> Define problem and inputs.
    [2] Secret sharing of inputs.
    [3] Compute secret shares.
    [4] Reconstruct the final result.
    [5] Output verification.
    [6] Decrypt the final result.
    </completion>
    <prompt>Substeps for: [2] Secret sharing of inputs.
    [2.1]</prompt><completion> Generate random secret shares.
    [2.2] Securely store secret shares.
    </completion>
    <prompt>Substeps for: [3] Compute secret shares. 
    [3.1]</prompt><completion> Encrypt secret share.
    [3.2] Distribute encrypted shares.
    [3.3] Compute ciphertext.
    [3.4] Broadcast result.</completion>
\end{lstlisting}

We combine the first-layer steps and some of the substeps in one \gls{aut} as we did in the ``dentist appointment" example. The \gls{aut} is shown in Figure \ref{fig: smpc}, where we indicate the first-layer steps and second-layer substeps in black and blue, respectively.
This example demonstrates the capability of the proposed approach to automatically distill domain-specific knowledge into control-oriented automata representations from natural-language task descriptions, even when applied to highly specialized problems.
We provide additional examples of domain-specific tasks in the Appendix.

\subsection{Demonstrations of Automaton Verification and Refinement}
After we use GLM2FSA to construct the automaton-based controller from queries to \GPT{}, we want to verify whether the knowledge encoded in the controller is consistent with the knowledge from other independent sources. To do this verification, we obtain a model and specifications from an independent knowledge source and use a model checker to verify whether the controller operating in the model satisfies the specification. 
We also present examples of the application of both of the procedures for iterative refinement that were described in Section \ref{sec:refinement}.

\paragraph{Crossing the Road Example: Manual Refinement Through Counterexamples.}

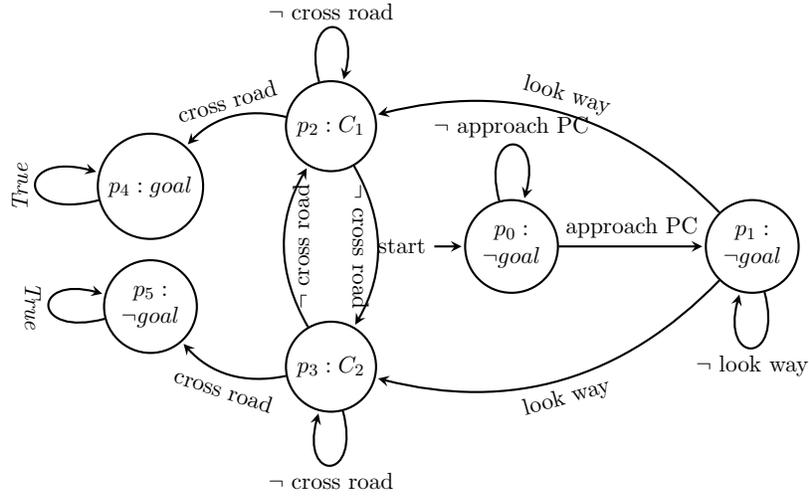
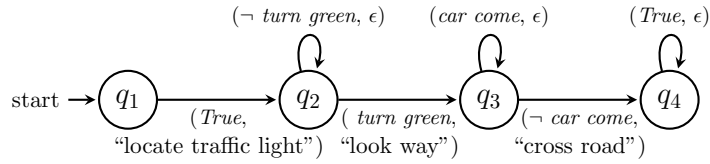
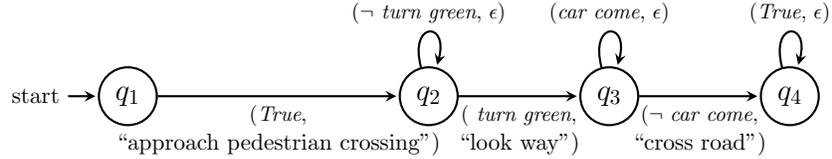
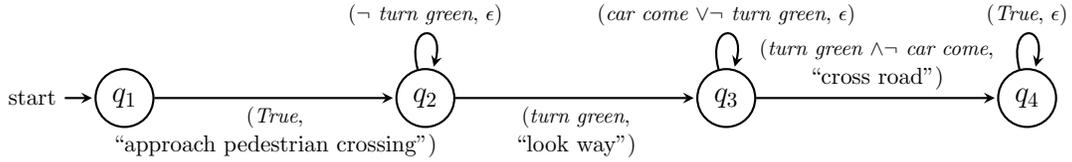
\begin{figure}[t]
    \begin{subfigure}{\linewidth}
        \centering
        \begin{tikzpicture}[
    scale=.8,
    node distance=2.2cm,
    thick,
    every node/.append style={transform shape},
]

\node[state,initial] (p0)
    at (0,0)
    {\shortstack{$p_0:$ \\ $\neg goal$}};
\node[state] (p1)
    at (4,0)
    {\shortstack{$p_1:$ \\ $\neg goal$}};
\node[state] (p2)
    at (-3,2)
    {$p_2:C_1$};
\node[state] (p3)
    at (-3,-2)
    {$p_3:C_2$};
\node[state] (p4)
    at (-6,1)
    {$p_4:goal$};
\node[state] (p5)
    at (-6,-1)
    {\shortstack{$p_5:$ \\ $\neg goal$}};

\path[->,sloped]

(p0) 
edge[loop above] node[sloped=false]
    {$\lnot$ approach PC}
    (p0)
edge[] node[]
    { approach PC }
    (p1)

(p1) 
edge[loop below] node[sloped=true]
    {$\lnot$ look way}
    (p1)
edge[bend right] node[]
    { look way }
    (p2)
edge[bend left] node[below]
    { look way }
    (p3)

(p2) 
edge[loop above] node[sloped=true, above]
    {$\neg$ cross road}
    (p2)
edge[bend left] node[below]
    { $\neg$ cross road}
    (p3)
edge[bend right] node[]
    { cross road }
    (p4)

(p3) 
edge[loop below] node[sloped=true]
    {$\neg$ cross road }
    (p3)
edge[bend left] node[below]
    {  $\neg$ cross road }
    (p2)
edge[bend left] node[below]
    {  cross road }
    (p5)

(p4) 
edge[loop left] node[above]
    { \textit{True} }
    (p4)

(p5) 
edge[loop left] node[below]
    { \textit{True} }
    (p5)

;

\end{tikzpicture}
        \caption{A model $\Aut[model]$ for ``cross the road." PC stands for Pedestrian Crossing, $C_1 =$ green $\land \neg$ car come $\land \neg$ goal, and $C_2 =$ (car come $\vee \neg$ green) $\land \neg$ goal. We annotate each node as ``state: label."
        }
        \label{fig: cross-verification}
    \end{subfigure}
    
    \vspace{1em}
    
    \begin{subfigure}{\linewidth}
        \centering
        \begin{tikzpicture}[thick,scale=.8, node distance=2.2cm, every node/.style={transform shape}]
	\node[state,initial] (1) at (0, 0) {\Large $q_1$};
	\node[state] (2) at (3, 0) {\Large $q_2$};
	\node[state] (3) at (6, 0) {\Large $q_3$};
    \node[state] (4) at (9, 0) {\Large $q_4$};
    
	\draw[->, shorten >=1pt] (1) to[left] node[below, align=center] {\small (\textit{True}, \\ ``locate traffic light")} (2);

    \draw[->, shorten >=1pt] (2) to[loop above] node[align=center] {\small (\textit{$\neg$ turn green}, $\noop$)} ();
 
	\draw[->, shorten >=1pt] (2) to[left] node[below, align=center, sloped] {\small ( \textit{turn green}, \\ ``look way")} (3);
	
    \draw[->, shorten >=1pt] (3) to[loop above] node[align=center] {\small (\textit{car come}, $\noop$)} ();
    \draw[->, shorten >=1pt] (3) to[left] node[below, align=center, sloped] {\small ($\neg$ \textit{car come}, \\ ``cross road")} (4);

    \draw[->, shorten >=1pt] (4) to[loop above] node[align=center] {\small (\textit{True}, $\noop$)} ();
\end{tikzpicture}
        \caption{
        The \gls{aut} for the task ``cross the road" under the condition \textit{at the traffic light = True}. We present only the bottom half of the controller in Figure \ref{fig: cross} for clarity.
        }
        \label{fig: cross-manual-1}
    \end{subfigure}
    
    \vspace{1em}
    
    \begin{subfigure}{\linewidth}
        \centering
        \begin{tikzpicture}[thick,scale=.8, node distance=2.2cm, every node/.style={transform shape}]
	\node[state,initial] (1) at (-2, 0) {\Large $q_1$};
	\node[state] (2) at (3, 0) {\Large $q_2$};
	\node[state] (3) at (6, 0) {\Large $q_3$};
    \node[state] (4) at (9, 0) {\Large $q_4$};
    
	\draw[->, shorten >=1pt] (1) to[left] node[below, align=center] {\small (\textit{True}, \\ ``approach pedestrian crossing")} (2);

    \draw[->, shorten >=1pt] (2) to[loop above] node[align=center] {\small (\textit{$\neg$ turn green}, $\noop$)} ();
 
	\draw[->, shorten >=1pt] (2) to[left] node[below, align=center, sloped] {\small ( \textit{turn green}, \\ ``look way")} (3);
	
    \draw[->, shorten >=1pt] (3) to[loop above] node[align=center] {\small (\textit{car come}, $\noop$)} ();
    \draw[->, shorten >=1pt] (3) to[left] node[below, align=center, sloped] {\small ($\neg$ \textit{car come}, \\ ``cross road")} (4);

    \draw[->, shorten >=1pt] (4) to[loop above] node[align=center] {\small (\textit{True}, $\noop$)} ();
\end{tikzpicture}
        \caption{The \gls{aut} after the first iteration of refinement.}
        \label{fig: cross-manual-2}
    \end{subfigure}
    
    \begin{subfigure}{\linewidth}
        \centering
        \begin{tikzpicture}[thick,scale=.8, node distance=2.2cm, every node/.style={transform shape}]
	\node[state,initial] (1) at (-2, 0) {\Large $q_1$};
	\node[state] (2) at (3, 0) {\Large $q_2$};
	\node[state] (3) at (8, 0) {\Large $q_3$};
    \node[state] (4) at (13, 0) {\Large $q_4$};
    
	\draw[->, shorten >=1pt] (1) to[left] node[below, align=center] {\small (\textit{True}, \\ ``approach pedestrian crossing")} (2);

    \draw[->, shorten >=1pt] (2) to[loop above] node[align=center] {\small (\textit{$\neg$ turn green}, $\noop$)} ();
 
	\draw[->, shorten >=1pt] (2) to[left] node[below, align=center, sloped] {\small (\textit{turn green}, \\ ``look way")} (3);
	
    \draw[->, shorten >=1pt] (3) to[loop above] node[align=center] {\small (\textit{car come $\vee \neg$ turn green}, $\noop$)} ();
    \draw[->, shorten >=1pt] (3) to[left] node[above, align=center, sloped] {\small (\textit{turn green $\land \neg$ car come}, \\ ``cross road")} (4);

    \draw[->, shorten >=1pt] (4) to[loop above] node[align=center] {\small (\textit{True}, $\noop$)} ();
\end{tikzpicture}
        \caption{The final \gls{aut} after two iterations of refinement.}
        \label{fig: cross-manual-3}
    \end{subfigure}
    
    \caption{The iterations of refinements. The user manually queries \GPT{} to refine the steps based on the counterexample from the verification.}
\end{figure}

We begin by verifying the correctness of the controller $\Aut[controller]$ from Figure \ref{fig: cross-manual-1} using the model $\Aut[model]$ illustrated in Figure \ref{fig: cross-verification}.
The specification is 
\begin{center}
    $\fLTL[spec] = \text{ traffic light} \land \lalways \leventually (\text{ green $\land \neg$ car come} ) \limplies \leventually \mathop{goal}$ 
\end{center}
(if the agent is at a traffic light and there always will eventually be a time when the traffic light is green and no car is coming, then it should eventually reach the goal).

The refinement of the controller in this example takes three iterations. In the first iteration, the verification step fails, and the model checker returns a counterexample that consists of an infinite loop of states $p_0, p_0, \ldots$, where $p_0$ is the initial state of the model from Figure \ref{fig: cross-verification}. 
We note that the counterexample is a sequence of states $(p_0, q_1), (p_0, q_2), (p_0, q_3), (p_0, q_3), ...$ from the product automaton.
However, to simplify the presentation, in the following exposition we only present the states $p_0, p_0, ...$ from the model when listing the counterexamples that were observed in experiments.

The trajectory of the infinite loop is provided to the user, consisting of the controller input symbol [$\neg$ approach pedestrian crossing] and output symbol [$\neg$ goal], repeated indefinitely. In words, the counterexample tells us that the controller fails because the controller never takes the action ``approach pedestrian crossing", which is required by the model.
In this example, the model-checking procedure thus automatically catches a missing action in the controller generated by \GLMtoDFA{}, which is provided through the external task knowledge (the model).
The user can then modify the input prompt to the language model to refine the controller as follows:
\begin{lstlisting}[language=completion]
    <prompt>Refine the following steps with an action "approach pedestrian crossing":
    [1] Locate the traffic light.
    ...
    [4] Cross the road if no cars are coming.
    </prompt>
    <completion>
    [1] Approach the pedestrian crossing.
    [2] Wait for the traffic light to turn green.
    [3] Look both ways before crossing the road.
    [4] Cross the road if no cars are coming.</completion>
\end{lstlisting}

The algorithm \GLMtoDFA{} then reconstructs the controller while including the previously missing action. We present the controller after the first iteration of refinement in Figure \ref{fig: cross-manual-2}. Then, the algorithm starts the next iteration.

In the second iteration, the refined controller appears correct with respect to the task and specification: It specifies that the agent should approach the pedestrian crossing, wait for the light to turn green, look both ways, wait for all cars to stop coming through the intersection, and cross the road.
However, the refined controller fails the verification step again with a counterexample $p_0 \rightarrow p_1 \rightarrow p_3 \rightarrow [\text{infinite loop } p_5]$.
State $p_{5}$ in $\Aut[model]$ is reached because it is possible for the controller to take the action ``cross road" when the traffic light is red.
This mistake could happen in scenarios where the traffic light switches from green to red while the agent is waiting for cars to stop coming.
This is a potentially dangerous edge case that the \gls{glm} fails to consider.
We emphasize that this edge case could easily be missed by a human as well. 
It is only by formally verifying the possible behaviors of the system against the model that the potential problem becomes apparent.

To handle the above corner case, we simply need to ensure that the traffic light is green and that there are simultaneously no cars coming before taking action ``cross road.''
So, to address the issue, the user can modify the input prompt as follows:

\begin{lstlisting}[language=completion]
    <prompt>Refine the following steps to ensure the action "cross the road" is performed under conditions "traffic light turns green" and "no cars are coming":
    [1] Approach the pedestrian crossing.
    ...
    [4] Cross the road if no cars are coming.
    </prompt>
    <completion>
    [1] Approach the pedestrian crossing.
    [2] Wait for the traffic light to turn green.
    [3] Look both ways before crossing the road.
    [4] Cross the road if no cars are coming and the traffic light is green.</completion>
\end{lstlisting}

We present the refined controller after the second iteration in Figure \ref{fig: cross-manual-3}. In the third iteration, the controller passes all the verification steps and hence it is finalized.

\paragraph{Crossing the Road Example: Automatic Refinement.}

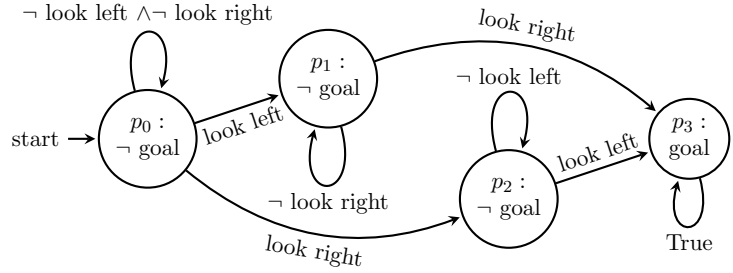
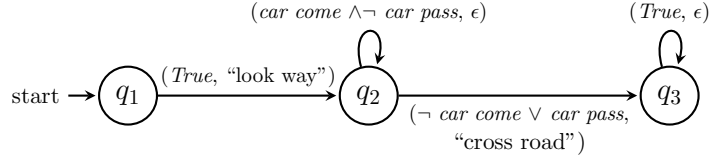
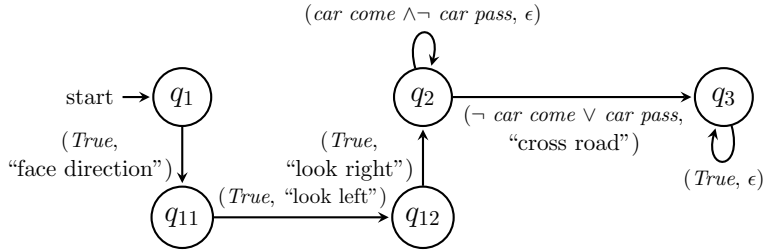
\begin{figure}[t]
\begin{subfigure}{\linewidth}
    \centering
    \begin{tikzpicture}[
    scale=.8,
    node distance=2.2cm,
    thick,
    every node/.append style={transform shape},
]

\node[state,initial] (p0)
    at (0,0)
    {\shortstack{$p_0:$ \\ $\neg$ goal }};
\node[state] (p1)
    at (3,1)
    {\shortstack{$p_1:$ \\ $\neg$ goal }};
\node[state] (p2)
    at (6,-1)
    {\shortstack{$p_2:$ \\ $\neg$ goal }};
\node[state] (p3)
    at (9,0)
    {\shortstack{$p_3:$ \\ goal }};

\path[->,sloped]

(p0) 
edge[loop above] node[sloped=false]
    {$\lnot$ look left $\land \lnot$ look right }
    (p0)
edge[] node[below]
    { look left }
    (p1)
edge[bend right] node[below]
    { look right }
    (p2)

(p1) 
edge[bend left] node[]
    { look right }
    (p3)
edge[loop below] node[sloped=false]
    { $\lnot$ look right }
    ()
    
(p2) 
edge[] node[]
    { look left }
    (p3)
edge[loop above] node[sloped=false]
    { $\lnot$ look left }
    ()

(p3) 
edge[loop below] node[]
    { True }
    ()

;

\end{tikzpicture}
    \caption{A model $\Aut[model]$ used to verify the \gls{aut} for the first-layer steps of the task ``cross the road."
    }
    \label{fig: cross-verification2}
\end{subfigure}

\vspace{1em}

\begin{subfigure}{\linewidth}
    \centering
    \begin{tikzpicture}[thick,scale=.8, node distance=2.2cm, every node/.style={transform shape}]
	\node[state,initial] (1) at (0, 0) {\Large $q_1$};
	\node[state] (2) at (4, 0) {\Large $q_2$};
	\node[state] (3) at (9, 0) {\Large $q_3$};
    
	\draw[->, shorten >=1pt] (1) to[left] node[above, align=center] {\small (\textit{True}, ``look way")} (2);
 
	\draw[->, shorten >=1pt] (2) to[left] node[below, align=center, sloped] {\small ($\neg$ \textit{car come $\vee$ car pass}, \\ ``cross road")} (3);
	
	\draw[->, shorten >=1pt] (2) to[loop above] node[align=center] {\small (\textit{car come $\land \neg$ car pass}, $\noop$)} ();

    \draw[->, shorten >=1pt] (3) to[loop above] node[align=center] {\small (\textit{True}, $\noop$)} ();
\end{tikzpicture}
    \caption{
    The \gls{aut} for the first-layer steps of the task ``cross the road" under the condition \textit{at traffic light = False}. We present the top half of the controller in Figure \ref{fig: cross} for clarity.
    }
    \label{fig: cross1}
\end{subfigure}

\vspace{1em}

\begin{subfigure}{\linewidth}
    \centering
    \begin{tikzpicture}[thick,scale=.8, node distance=2.2cm, every node/.style={transform shape}]
	\node[state,initial] (1) at (0, 2) {\Large $q_1$};
	\node[state] (11) at (0, 0) {\Large $q_{11}$};
    \node[state] (12) at (4, 0) {\Large $q_{12}$};
    \node[state] (2) at (4, 2) {\Large $q_2$};
	\node[state] (3) at (9, 2) {\Large $q_3$};
    
	\draw[->, shorten >=1pt] (1) to[left] node[align=center] {\small (\textit{True}, \\ ``face direction")} (11);

    \draw[->, shorten >=1pt] (11) to[left] node[above, align=center] {\small (\textit{True}, ``look left")} (12);

    \draw[->, shorten >=1pt] (12) to[left] node[align=center] {\small (\textit{True}, \\ ``look right")} (2);
 
	\draw[->, shorten >=1pt] (2) to[left] node[below, align=center, sloped] {\small ($\neg$ \textit{car come $\vee$ car pass}, \\ ``cross road")} (3);
	
	\draw[->, shorten >=1pt] (2) to[loop above] node[align=center] {\small (\textit{car come $\land \neg$ car pass}, $\noop$)} ();

     \draw[->, shorten >=1pt] (3) to[loop below] node[align=center] {\small (\textit{True}, $\noop$)} ();
\end{tikzpicture}
    \caption{The refined \gls{aut}.}
    \label{fig: cross2}
\end{subfigure}

\caption{The model $\Aut[model]$ used to verify and refine the substeps in the task ``cross the road."}
\end{figure}

Figure \ref{fig: cross-verification2} shows a model that we use to verify the ability of the controller in Figure \ref{fig: cross} to look left and right. 
We impose the specification $\fLTL[spec] = \lnot \text{ traffic light} \limplies \leventually \mathop{goal}$.
However, under this model, the controller from Figure \ref{fig: cross1} fails to satisfy the specification.
Therefore, we follow the refinement procedure and query \GPT{} for second-layer steps (substeps of the first-layer steps) and we construct a new controller to represent these second-layer steps. 
The resulting controller is presented in Figure \ref{fig: substep-cross}.
Next, we remove the states following the pruning process described above, which results in the controller illustrated in Figure \ref{fig: cross2}. 

We highlight the fact that the pruned controller in Figure \ref{fig: cross2} satisfies the specification shown in Figure \ref{fig: cross-verification2}. 
So, the refinement procedure terminates, and we do not send further queries to \GPT{}. 
Furthermore, because we have pruned the unnecessary steps, this new controller also resolves the logical flaw raised in our first examination of this example.

\paragraph{WiFi Troubleshooting: Verb Phrase Matching and Manual Refinement.}
We construct a controller for rebooting the modem and router to solve the internet connection problem. 
We first query \GPT{} on the steps required to reboot both devices:
\begin{lstlisting}[language=completion]
    <prompt>There is an internet connection problem and I need to reboot both the modem and router.
    Steps for rebooting the modem and router:
    [1]</prompt><completion> Unplug the modem's power cord
    [2] Disconnect the router's power source
    [3] Reconnect the modem's power cord
    [4] Observe the modem's indicator lights
    [5] Reconnect the router's power source
    [6] Monitor the router's indicator lights
    [7] Confirm internet connectivity on devices</completion>
\end{lstlisting}
We obtain a list of steps in natural language from \GPT{} and following \GLMtoDFA{} to construct an automaton-based controller, as presented in Figure \ref{fig: wifi}. The transitions in black are the transitions of the controller we originally constructed using the responses from \GPT{}.

\begin{figure*}[t]
    \centering
    \begin{tikzpicture}[thick,scale=.7, node distance=2.2cm, every node/.style={transform shape}]
	\node[state,initial] (1) at (0, 2) {\Large $q_{1}$};
	\node[state] (2) at (4, 4) {\Large $q_{2}$};
    \node[state] (3) at (8, 4) {\Large $q_{3}$};
	\node[state] (4) at (12, 2) {\Large $q_{4}$};
    \node[state] (5) at (10, -2) {\Large $q_{5}$};
	\node[state] (6) at (6, -2) {\Large $q_6$};
    \node[state] (7) at (0, -1) {\Large $q_7$};
    \node[state] (8) at (-4, 0) {\Large $q_8$};

    \node[state, color=purple] (41) at (8, 2.5) {\Large $q_{4'}$};
    \node[state, color=brown] (61) at (3.5, 2.5) {\Large $q_{6'}$};

	\draw[->, shorten >=1pt, sloped] (1) to[bend left] node[above, align=center] {\small (\textit{True}, \\ ``unplug modem power")} (2);
    
	\draw[->, shorten >=1pt, sloped] (2) to[bend left] node[above, align=center] {\small (\textit{True}, \\ ``disconnect router power")} (3);
    
    \draw[->, shorten >=1pt, sloped] (3) to[bend left] node[above, align=center] {\small (\textit{True}, \\ ``reconnect modem power")} (4);
    \draw[->, shorten >=1pt, sloped, dashed] (4) to[bend left] node[below, align=center] {\small (\textit{True}, \\ ``observe modem indicator")} (5);
    \draw[->, shorten >=1pt, sloped] (5) to[bend left] node[below, align=center] {\small (\textit{True}, \\ ``reconnect router power")} (6);
    \draw[->, shorten >=1pt, sloped, dashed] (6) to[bend left] node[below, align=center] {\small (\textit{True}, \\ ``observe router indicator")} (7);

     \draw[->, shorten >=1pt, sloped] (7) to[left] node[below, align=center] {\small (\textit{True}, \\ ``confirm internet")} (8);

     \draw[->, shorten >=1pt] (8) to[loop above] node[align=center] {\small (\textit{True}, $\noop$)} ();

     \draw[->, shorten >=1pt, color=purple] (4) to[loop right] node[align=center] {\small ( $\neg$ \textit{2 min}, $\noop$)} ();
     \draw[->, shorten >=1pt, sloped, color=purple] (4) to[left] node[below, align=center] {\small (\textit{2 min}, $\noop$)} (41);
     \draw[->, shorten >=1pt, sloped, color=purple] (41) to[bend right] node[below, align=center] {\small (\textit{True}, \\ ``observe modem indicator")} (5);

     \draw[->, shorten >=1pt, color=brown] (6) to[loop above] node[align=center] {\small ( $\neg$ \textit{2 min}, \\ $\noop$)} ();
     \draw[->, shorten >=1pt, sloped, color=brown] (6) to[left] node[below, align=center] {\small (\textit{2 min}, $\noop$)} (61);
     \draw[->, shorten >=1pt, sloped, color=brown] (61) to[left] node[below, align=center] {\small (\textit{True}, \\ ``observe router indicator")} (7);
\end{tikzpicture}
    \caption{The \gls{aut}-based controller for fixing the internet connection. The transitions in black are from the original controller. After the first iteration of refinement, we replace the transition between $q_4$ and $q_5$ to the transitions in {\color{purple} purple}. After the second iteration of refinement, we replace the transition between $q_6$ and $q_7$ to the transitions in {\color{brown} brown}. The removed transitions are dashed and the final controller consists only of the transitions drawn with solid lines.}
    \label{fig: wifi}
\end{figure*}

\begin{figure}[t]
    \centering
    \begin{tikzpicture}[
    scale=.8,
    node distance=2.2cm,
    thick,
    every node/.append style={transform shape},
]

\node[state,initial] (q0)
    at (0,0)
    {\shortstack{$p_0:$ \\ $\neg$ goal }};
\node[state] (q1)
    at (4,2)
    {\shortstack{$p_1:$ \\ $\neg$ goal }};
\node[state] (q2)
    at (9,2)
    {\shortstack{$p_2:$ \\ $\neg$ 2 min}};
\node[state] (q3)
    at (12,1)
    {\shortstack{$p_3:$ \\ 2 min }};
\node[state] (q4)
    at (8,-2)
    {\shortstack{$p_4:$ \\ $\neg$ 2 min }};
\node[state] (q5)
    at (4,-2)
    {\shortstack{$p_5:$ \\ $\neg$ goal }};
\node[state] (q6)
    at (12,-2)
    {\shortstack{$p_6:$ \\ 2 min $\land$ goal }};

\path[->,sloped]

(q0) 
edge[loop right] node[sloped=false]
    {$\lnot$ unplug modem }
    ()
edge[bend left] node[]
    { unplug modem }
    (q1)

(q1) 
edge[] node[]
    { plug-in modem }
    (q2)
edge[loop above] node[]
    { $\lnot$ plug-in modem }
    ()

(q2) 
edge[bend left] node[]
    { $\noop$ }
    (q3)
edge[] node[]
    {$\lnot \noop$}
    (q5)

(q3) 
edge[loop right] node[above]
    { $\neg$ turn on router }
    ()
edge[] node[]
    { turn on router }
    (q4)

(q4) 
edge[] node[]
    { $\noop$ }
    (q6)
edge[] node[]
    {$\lnot \noop$}
    (q5)

(q5) 
edge[loop left] node[above]
    {$\ltrue$}
    (q5)

(q6) 
edge[loop right] node[below]
    { $\ltrue$ }
    ()

;

\end{tikzpicture}
    \caption{The model represents the side information available from an independent knowledge source. We use it to verify the controller in Figure \ref{fig: wifi} against specification $\Phi = \leventually goal$. }
    \label{fig: wifi-model}
\end{figure}
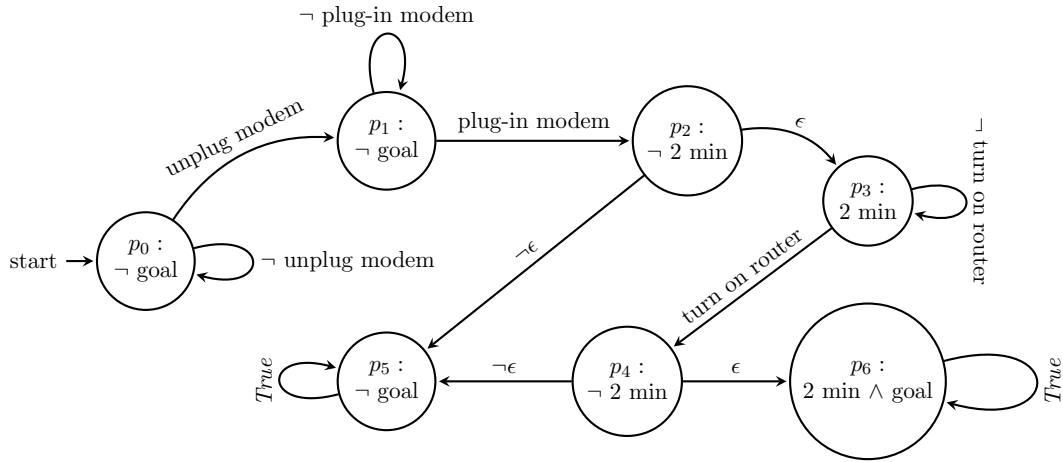

Then, a model encoding the information of rebooting the modem and router from Spectrum Service is provided, as presented in Figure \ref{fig: wifi-model}. We verify whether the knowledge from \GPT{} is consistent with the knowledge from Spectrum Service. Since the controller and model encode knowledge from different sources, the phrases they use are different. 
Therefore, we match the verb phrases from the controller and model before proceeding with the verification procedure, as described in Section \ref{sec:verification_subsection}.
\begin{lstlisting}[language=completion]
    <prompt>Do the following pairs of verb phrases lead to the same effect?
    1, "Unplug the modem's power cord" and "Unplug the modem"
    2, "Disconnect the router's power source" and "Turn off the router"
    3, "Reconnect the modem's power cord" and "Plug in the modem"
    4, "Reconnect the router's power source" and "Turn on the router"</prompt>
    <completion> 
    1. Yes, both phrases lead to cutting power to the modem.
    2. Yes, both phrases lead to cutting power to the router.
    3. Yes, both phrases lead to restoring power to the modem.
    4. Yes, both phrases lead to restoring power to the router.</completion>
\end{lstlisting}

Once the synonymous verb phrases have been consolidated, we verify that the controller satisfies the specification $\Phi = \leventually goal$ when implemented against the model.
In the first iteration, the verification fails and the model checker returns a counterexample of states $p_0 \rightarrow p_1 \rightarrow p_2 \rightarrow  [\text{infinite loop } p_5]$. 
Intuitively, this failure occurs because the initial controller generated by GLM2FSA does not wait for any amount of time before restarting the modem, however, the model derived from the information provided by Spectrum Service requires such a waiting period.
The user thus uses the following input prompt to \GPT{} to refine the controller:
\begin{lstlisting}[language=completion]
    <prompt>Revise the following steps to include "wait two minutes" after "plug in modem":
    [1] Unplug the modem's power cord
    ......
    [7] Confirm internet connectivity on devices
    [1]</prompt><completion> Unplug the modem's power cord
    [2] Disconnect the router's power source
    [3] Reconnect the modem's power cord
    [4] Wait two minutes
    [5] Observe the modem's indicator lights
    [6] Reconnect the router's power source
    [7] Monitor the router's indicator lights
    [8] Confirm internet connectivity on devices</completion>
\end{lstlisting}

We run the algorithm \GLMtoDFA{} again on the new responses from the language model to construct a new controller. Compared to the original controller, the {\color{purple} purple} transitions in Figure \ref{fig: wifi} are added and the original transition from $q_4$ to $q_5$ is removed.

In the second iteration, we run the model checker again on the refined controller and get another counterexample $p_0 \rightarrow p_1 \rightarrow p_2 \rightarrow p_3 \rightarrow p_4 \rightarrow [\text{infinite loop } p_5]$. The user again interprets the reason for the failure, refines the input prompt, and obtains a list of refined steps from \GPT{}:
\begin{lstlisting}[language=completion]
    <prompt>Revise the following steps to include "wait two minutes" after "turn on router":
    [1] Unplug the modem's power cord
    ......
    [8] Confirm internet connectivity on devices
    [1]</prompt><completion> Unplug the modem's power cord
    [2] Disconnect the router's power source
    [3] Reconnect the modem's power cord
    [4] Wait two minutes
    [5] Observe the modem's indicator lights
    [6] Reconnect the router's power source
    [7] Wait two minutes
    [8] Monitor the router's indicator lights
    [9] Confirm internet connectivity on devices</completion>
\end{lstlisting}
After the second iteration of refinement, the controller passes the model checking step. We present the controller in Figure \ref{fig: wifi}, in which the transitions in solid lines are the controller after refinements. 
\section{Conclusions}

We provide a proof-of-concept for the automatic construction of automaton-based representations of abstract task knowledge from \glspl{glm}.
We propose an algorithm, \GLMtoDFA{}, that accepts brief natural-language descriptions of tasks as input, queries a \gls{glm}, and then constructs an automaton from the language model's responses.
The algorithm is highly automated, requiring only a short task description to build machine-understandable knowledge representations.
We additionally propose methods to formally verify the automata output by \GLMtoDFA{}, and to use the results of verification to iteratively refine the inputs to the \gls{glm}.
Experimental results demonstrate the capabilities of \GLMtoDFA{}.
The generated automaton-based controllers capture domain-specific knowledge, even when applied to highly specialized problems. 
Furthermore, the procedure for verification-guided controller refinement can automatically catch, explain, and fix, the occasional nonsensical or incorrect behaviors that are output by the \gls{glm}. 




\appendix


\section{Additional Background and Definitions}
\label{sec:appendix_background}
\paragraph{\GlSentrylongpl{glm}.}
A \glsreset{glm}\gls{glm} produces human-like text \emph{completion} from a given initial text (\emph{prompt}).
The produced texts continue filling the content from that prompt.
Recent \glspl{glm} are deep learning models with millions or billions of parameters; hence they are called large-scale GLMs.

Generative Pre-trained Transformer 3 (\GPT{}) is the current state-of-the-art large-scale GLM.
It is pretrained by five large-scale datasets with over 5 billion words.
\GPT{} offers four primary models with different capabilities for different tasks \cite{brown2020GPT3}.
\emph{Davinci-003} is the most capable model that can do question-answering, next-sentence prediction, and text insertion.
We query \emph{Davinci-003} to obtain task instructions for empirical analysis.

\GPT{} allows users to customize settings by setting the hyper-parameters.
For instance, \emph{max\_tokens} restricts the maximum number of \emph{tokens} (words and punctuation) of the generated text, and \emph{temperature} defines the randomness of the outputs.
We propose an algorithm that specifies grammar rules for certain keywords.
Hence we set \emph{bias} on the keywords to ensure the model outputs them instead of their alternations.
Setting bias to keywords eliminates the need to transform synonyms to the corresponding keywords or define new rules for those synonyms.

\paragraph{\GlSentrylong{ltl}.} 
Formally, \gls{ltl} formulas are defined inductively as:
$
\varphi \coloneqq \Autprop\in\AutProps[model] \mid \lnot\varphi \mid \varphi \lor \varphi \mid \lnext\varphi \mid \varphi\luntil\varphi
$
Intuitively, an \gls{ltl} formula consists of
\begin{itemize}
    \item A set of atomic propositions, denoted by lowercase letters (e.g., car come), represent the system's state.
    \item A set of temporal operators describes the system's temporal behavior.
    \item A set of logical connectives, such as negation ($\neg$), conjunction ($\land$), and disjunction ($\vee$), that can be used to combine atomic propositions and temporal operators.
\end{itemize}

As syntax sugar, along with additional constants and operators used in propositional logic,
we allow the standard temporal operators $\leventually$ (``eventually'') and $\lalways$ (``always'').

\section{Verification using NuSMV~\cite{Cimatti2002NuSMV}.}

Here we provide the NuSMV implementation used to verify the controllers from \Cref{fig: cross,fig: substep-cross} under a model from \Cref{fig: cross-verification}.

\begin{lstlisting}[language=NuSMV]

MODULE environment -------------------------------------------------------------

VAR
    car_come: boolean;          -- cars are coming
    car_pass: boolean;          -- all cars passed
    turn_green: boolean;        -- the pedestrian traffic light turned green

FROZENVAR
    traffic_light: boolean;     -- there is a pedestrian traffic light

MODULE actions -----------------------------------------------------------------

VAR
    face_direction: boolean;    -- face direction of crossing
    look_left: boolean;         -- look left for cars
    look_right: boolean;        -- look right for cars
    look_way: boolean;          -- look both ways for cars
    cross: boolean;             -- cross the road

DEFINE COUNT_ACTIONS := count(
    face_direction,
    look_left,
    look_right,
    cross,
    look_way
);
DEFINE none := COUNT_ACTIONS = 0;   -- true iff no action is taken

INVAR COUNT_ACTIONS <= 1; -- cannot take two actions at once
INIT none;


MODULE controller_fig4(env,act) ------------------------------------------------

VAR state: {
    0,
    11,12,
    21,22,
    3
};
INIT            -- the initial state
    state=0;
DEFINE goal :=  -- the desired state
    state=3;

TRANS
    case
        state=0 & next(!env.traffic_light)
        : next(act.none & state=11);
        state=0 & next(env.traffic_light)
        : next(act.none & state=21);

        (state=11)
        : next(act.look_way & state=12);

        state=12 & next(env.car_come & !env.car_pass)
        : next(act.none & state=12);
        state=12 & next(!env.car_come | env.car_pass)
        : next(act.cross & state=3);

        (state=21)
        : next(act.look_way & state=22);

        state=22 & next(!env.turn_green)
        : next(act.none & state=12);
        state=22 & next(env.turn_green)
        : next(act.cross & state=3);

        goal
        : next(act.none & goal);
    esac;

MODULE controller_fig5(env,act) ------------------------------------------------

VAR state: {
    11,12,13,14,
    21,22,23,
    31,32,
    4
};
INIT            -- the initial state
    state=11;
DEFINE goal :=  -- the desired state
    state=4;

TRANS
    case
        state=11
        : next(act.face_direction & state=12);

        state=12
        : next(act.look_left & state=13);

        state=13
        : next(act.look_right & state=14);

        state=14 & next(!env.car_come)
        : next(act.none & state=21);
        state=14 & next(env.car_come)
        : next(act.none & state=31);

        state=21
        : next(act.cross & state=22);

        state=22
        : next(act.look_way & state=23);

        state=23 & next(!env.car_come)
        : next(act.none & state=4);
        state=23 & next(env.car_come)
        : next(act.none & state=11);

        state=31 & next(!env.car_pass)
        : next(act.none & state=31);
        state=31 & next(env.car_pass)
        : next(act.none & state=32);

        state=32
        : next(act.none & state=21);

        state=4
        : next(act.none & state=4);
    esac;


MODULE model(env,act) ----------------------------------------------------------

VAR state: {
    q_init,q_sink,q_goal
};
INIT            -- the initial state
    state=q_init;
DEFINE goal :=  -- the desired state
    state=q_goal;

TRANS
    case
        state=q_init & next(!act.cross)
        : next(state=q_init);
        state=q_init & next(act.cross & env.car_come)
        : next(state=q_sink);
        state=q_init & next(act.cross & !env.car_come)
        : next(state=q_goal);

        state=q_sink
        : next(state=q_sink);
        
        state=q_goal
        : next(state=q_goal);
    esac;

MODULE main --------------------------------------------------------------------

VAR env: environment;
VAR act: actions;

VAR controller: controller_fig4(env,act);
-- VAR controller: controller_fig5(env,act);
VAR model: model(env,act);

LTLSPEC NAME assume_guarantee :=
    (  TRUE -- ASSUMPTIONS

        -- the flow of cars is not continuous
        &  (G F !env.car_come)

        -- cars coming eventually passes
        &  (G (env.car_come -> F env.car_pass))

        -- if all cars passed,
        -- then none are coming towards the crossing
        &  (G (env.car_pass -> !env.car_come))

        -- -- if all cars passed,
        -- -- then none are coming towards the crossing for a bit
        -- &  (G (env.car_pass -> G[0,2] !env.car_come))

        -- -- if one has looked left and right recently and no car is coming,
        -- -- no car will be coming for a bit
        -- &  (G (
        --     (!env.car_come & (O[0,2] act.look_left) & (O[0,2] act.look_right))
        --     -> (G[0,1] !env.car_come)
        -- ))

        -- when pedestrian light is green, no car can come on the crossing
        &  (G ((env.traffic_light & env.turn_green) -> (!env.car_come)))

    ) -> ( TRUE -- GUARANTEES

        -- the model terminal state is eventually reached
        &  (F model.goal)

        -- eventually, the controller rightfully thinks it achieved its goal
        &  (F (model.goal & controller.goal))

    );
\end{lstlisting}

\bibliography{references}

\begin{thebibliography}{}

\bibitem[\protect\BCAY{Arnaiz-Gonz{\'a}lez, D{\'\i}ez-Pastor, Ramos-P{\'e}rez,\
  \BBA\ Garc{\'\i}a-Osorio}{Arnaiz-Gonz{\'a}lez
  et~al.}{2018}]{ArnaizGonzlez2018SeshatA}
Arnaiz-Gonz{\'a}lez, {\'A}., D{\'\i}ez-Pastor, J.-F., Ramos-P{\'e}rez, I.,
  \BBA\ Garc{\'\i}a-Osorio, C. \BBOP2018\BBCP.
\newblock \BBOQ Seshat—a web-based educational resource for teaching the most
  common algorithms of lexical analysis\BBCQ\
\newblock {\Bem Computer Applications in Engineering Education}, {\Bem
  26\/}(6), 2255--2265.

\bibitem[\protect\BCAY{Baier\ \BBA\ Katoen}{Baier\ \BBA\
  Katoen}{2008}]{baier2008principles}
Baier, C.\BBACOMMA\  \BBA\ Katoen, J.-P. \BBOP2008\BBCP.
\newblock {\Bem Principles of Model Checking}.
\newblock MIT Press.

\bibitem[\protect\BCAY{Baral, Dzifcak, Gonzalez,\ \BBA\ Zhou}{Baral
  et~al.}{2011}]{Baral2011UsingIL}
Baral, C., Dzifcak, J., Gonzalez, M.~A., \BBA\ Zhou, J. \BBOP2011\BBCP.
\newblock \BBOQ Using inverse lambda and generalization to translate english to
  formal languages\BBCQ\
\newblock In {\Bem International Conference on Computational Semantics}, \BPGS\
  35--44.

\bibitem[\protect\BCAY{Brouwer, Gellerich,\ \BBA\ Pl{\"{o}}dereder}{Brouwer
  et~al.}{1998}]{Brouwer1998MythsAF}
Brouwer, K., Gellerich, W., \BBA\ Pl{\"{o}}dereder, E. \BBOP1998\BBCP.
\newblock \BBOQ Myths and facts about the efficient implementation of finite
  automata and lexical analysis\BBCQ\
\newblock In {\Bem Compiler Construction}, \lowercase{\BVOL}\ 1383 of {\Bem
  Lecture Notes in Computer Science}, \BPGS\ 1--15.

\bibitem[\protect\BCAY{Brown, Mann, Ryder, Subbiah, Kaplan, Dhariwal,
  Neelakantan, Shyam, Sastry, Askell, Agarwal, Herbert{-}Voss, Krueger,
  Henighan, Child, Ramesh, Ziegler, Wu, Winter, Hesse, Chen, Sigler, Litwin,
  Gray, Chess, Clark, Berner, McCandlish, Radford, Sutskever,\ \BBA\
  Amodei}{Brown et~al.}{2020}]{brown2020GPT3}
Brown, T.~B., Mann, B., Ryder, N., Subbiah, M., Kaplan, J., Dhariwal, P.,
  Neelakantan, A., Shyam, P., Sastry, G., Askell, A., Agarwal, S.,
  Herbert{-}Voss, A., Krueger, G., Henighan, T., Child, R., Ramesh, A.,
  Ziegler, D.~M., Wu, J., Winter, C., Hesse, C., Chen, M., Sigler, E., Litwin,
  M., Gray, S., Chess, B., Clark, J., Berner, C., McCandlish, S., Radford, A.,
  Sutskever, I., \BBA\ Amodei, D. \BBOP2020\BBCP.
\newblock \BBOQ Language models are few-shot learners\BBCQ\
\newblock {\Bem Advances in Neural Information Processing Systems}, {\Bem 33},
  1877--1901.

\bibitem[\protect\BCAY{Cimatti, Clarke, Giunchiglia, Giunchiglia, Pistore,
  Roveri, Sebastiani,\ \BBA\ Tacchella}{Cimatti
  et~al.}{2002}]{Cimatti2002NuSMV}
Cimatti, A., Clarke, E.~M., Giunchiglia, E., Giunchiglia, F., Pistore, M.,
  Roveri, M., Sebastiani, R., \BBA\ Tacchella, A. \BBOP2002\BBCP.
\newblock \BBOQ Nu{SMV} 2: An opensource tool for symbolic model checking\BBCQ\
\newblock In {\Bem Computer Aided Verification}, \lowercase{\BVOL}\ 2404 of
  {\Bem Lecture Notes in Computer Science}, \BPGS\ 359--364.

\bibitem[\protect\BCAY{Davison, Feldman,\ \BBA\ Rush}{Davison
  et~al.}{2019}]{davison2019commonsense}
Davison, J., Feldman, J., \BBA\ Rush, A.~M. \BBOP2019\BBCP.
\newblock \BBOQ Commonsense knowledge mining from pretrained models\BBCQ\
\newblock In {\Bem Conference on Empirical Methods in Natural Language
  Processing}, \BPGS\ 1173--1178.

\bibitem[\protect\BCAY{Fang, Wang, Yin, Han,\ \BBA\ Zhao}{Fang
  et~al.}{2020}]{Fang2020MultiagentRL}
Fang, X., Wang, J., Yin, C., Han, Y., \BBA\ Zhao, Q. \BBOP2020\BBCP.
\newblock \BBOQ Multiagent reinforcement learning with learning automata for
  microgrid energy management and decision optimization\BBCQ\
\newblock In {\Bem Chinese Control and Decision Conference}, \BPGS\ 779--784.

\bibitem[\protect\BCAY{Francis, Kitamura, Labelle, Lu, Navarro,\ \BBA\
  Oh}{Francis et~al.}{2022}]{evlp}
Francis, J., Kitamura, N., Labelle, F., Lu, X., Navarro, I., \BBA\ Oh, J.
  \BBOP2022\BBCP.
\newblock \BBOQ Core challenges in embodied vision-language planning\BBCQ\
\newblock {\Bem J. Artif. Intell. Res.}, {\Bem 74}, 459--515.

\bibitem[\protect\BCAY{Ghosh, Elenius, Li, Lincoln, Shankar,\ \BBA\
  Steiner}{Ghosh et~al.}{2016}]{Ghosh2014ARSENALAR}
Ghosh, S., Elenius, D., Li, W., Lincoln, P., Shankar, N., \BBA\ Steiner, W.
  \BBOP2016\BBCP.
\newblock \BBOQ {ARSENAL:} automatic requirements specification extraction from
  natural language\BBCQ\
\newblock In {\Bem {NASA} Formal Methods}, \lowercase{\BVOL}\ 9690 of {\Bem
  Lecture Notes in Computer Science}, \BPGS\ 41--46.

\bibitem[\protect\BCAY{He, Fang, Wang,\ \BBA\ Song}{He
  et~al.}{2022}]{He2022AcquiringAM}
He, M., Fang, T., Wang, W., \BBA\ Song, Y. \BBOP2022\BBCP.
\newblock \BBOQ Acquiring and modelling abstract commonsense knowledge via
  conceptualization\BBCQ.

\bibitem[\protect\BCAY{Hendrycks, Burns, Basart, Zou, Mazeika, Song,\ \BBA\
  Steinhardt}{Hendrycks et~al.}{2021}]{hendrycks2020measuring}
Hendrycks, D., Burns, C., Basart, S., Zou, A., Mazeika, M., Song, D., \BBA\
  Steinhardt, J. \BBOP2021\BBCP.
\newblock \BBOQ Measuring massive multitask language understanding\BBCQ\
\newblock In {\Bem International Conference on Learning Representations}.

\bibitem[\protect\BCAY{Honnibal, Montani, Van~Landeghem,\ \BBA\ Boyd}{Honnibal
  et~al.}{2020}]{spaCy}
Honnibal, M., Montani, I., Van~Landeghem, S., \BBA\ Boyd, A. \BBOP2020\BBCP.
\newblock \BBOQ spacy: Industrial-strength natural language processing in
  python\BBCQ.

\bibitem[\protect\BCAY{Huang, Abbeel, Pathak,\ \BBA\ Mordatch}{Huang
  et~al.}{2022a}]{huang2022language}
Huang, W., Abbeel, P., Pathak, D., \BBA\ Mordatch, I. \BBOP2022a\BBCP.
\newblock \BBOQ Language models as zero-shot planners: Extracting actionable
  knowledge for embodied agents\BBCQ\
\newblock In {\Bem International Conference on Machine Learning},
  \lowercase{\BVOL}\ 162 of {\Bem Proceedings of Machine Learning Research},
  \BPGS\ 9118--9147.

\bibitem[\protect\BCAY{Huang, Xia, Xiao, Chan, Liang, Florence, Zeng, Tompson,
  Mordatch, Chebotar, Sermanet, Jackson, Brown, Luu, Levine, Hausman,\ \BBA\
  Ichter}{Huang et~al.}{2022b}]{huang2022inner}
Huang, W., Xia, F., Xiao, T., Chan, H., Liang, J., Florence, P., Zeng, A.,
  Tompson, J., Mordatch, I., Chebotar, Y., Sermanet, P., Jackson, T., Brown,
  N., Luu, L., Levine, S., Hausman, K., \BBA\ Ichter, B. \BBOP2022b\BBCP.
\newblock \BBOQ Inner monologue: Embodied reasoning through planning with
  language models\BBCQ\
\newblock In {\Bem Conference on Robot Learning}, \lowercase{\BVOL}\ 205 of
  {\Bem Proceedings of Machine Learning Research}, \BPGS\ 1769--1782.

\bibitem[\protect\BCAY{Icarte, Klassen, Valenzano,\ \BBA\ McIlraith}{Icarte
  et~al.}{2018}]{icarte}
Icarte, R.~T., Klassen, T.~Q., Valenzano, R.~A., \BBA\ McIlraith, S.~A.
  \BBOP2018\BBCP.
\newblock \BBOQ Using reward machines for high-level task specification and
  decomposition in reinforcement learning\BBCQ\
\newblock In Dy, J.~G.\BBACOMMA\  \BBA\ Krause, A.\BEDS, {\Bem International
  Conference on Machine Learning}, \lowercase{\BVOL}~80 of {\Bem Proceedings of
  Machine Learning Research}, \BPGS\ 2112--2121.

\bibitem[\protect\BCAY{Ichter, Brohan, Chebotar, Finn, Hausman, Herzog, Ho,
  Ibarz, Irpan, Jang, Julian, Kalashnikov, Levine, Lu, Parada, Rao, Sermanet,
  Toshev, Vanhoucke, Xia, Xiao, Xu, Yan, Brown, Ahn, Cortes, Sievers, Tan, Xu,
  Reyes, Rettinghouse, Quiambao, Pastor, Luu, Lee, Kuang, Jesmonth, Joshi,
  Jeffrey, Ruano, Hsu, Gopalakrishnan, David, Zeng,\ \BBA\ Fu}{Ichter
  et~al.}{2022}]{brohan2023can}
Ichter, B., Brohan, A., Chebotar, Y., Finn, C., Hausman, K., Herzog, A., Ho,
  D., Ibarz, J., Irpan, A., Jang, E., Julian, R., Kalashnikov, D., Levine, S.,
  Lu, Y., Parada, C., Rao, K., Sermanet, P., Toshev, A., Vanhoucke, V., Xia,
  F., Xiao, T., Xu, P., Yan, M., Brown, N., Ahn, M., Cortes, O., Sievers, N.,
  Tan, C., Xu, S., Reyes, D., Rettinghouse, J., Quiambao, J., Pastor, P., Luu,
  L., Lee, K., Kuang, Y., Jesmonth, S., Joshi, N.~J., Jeffrey, K., Ruano,
  R.~J., Hsu, J., Gopalakrishnan, K., David, B., Zeng, A., \BBA\ Fu, C.~K.
  \BBOP2022\BBCP.
\newblock \BBOQ Do as {I} can, not as {I} say: Grounding language in robotic
  affordances\BBCQ\
\newblock In {\Bem Conference on Robot Learning}, \lowercase{\BVOL}\ 205 of
  {\Bem Proceedings of Machine Learning Research}, \BPGS\ 287--318.

\bibitem[\protect\BCAY{Kate, Wong,\ \BBA\ Mooney}{Kate
  et~al.}{2005}]{kate2005learning}
Kate, R.~J., Wong, Y.~W., \BBA\ Mooney, R.~J. \BBOP2005\BBCP.
\newblock \BBOQ Learning to transform natural to formal languages\BBCQ\
\newblock In {\Bem National Conference on Artificial Intelligence}, \BPGS\
  1062--1068.

\bibitem[\protect\BCAY{Kucera, Francis, Twaddell, Marckworth, Bell,\ \BBA\
  Carroll}{Kucera et~al.}{1967}]{Kucera1967ComputationalAO}
Kucera, H., Francis, W.~N., Twaddell, W.~F., Marckworth, M.~L., Bell, L.~M.,
  \BBA\ Carroll, J.~B. \BBOP1967\BBCP.
\newblock \BBOQ Computational analysis of present-day american english\BBCQ\
\newblock {\Bem International Journal of American Linguistics}, {\Bem 35},
  71--75.

\bibitem[\protect\BCAY{Lu, Feng, Zhu, Xu, Wang, Eckstein,\ \BBA\ Wang}{Lu
  et~al.}{2022}]{Lu2022NeuroSymbolicCL}
Lu, Y., Feng, W., Zhu, W., Xu, W., Wang, X.~E., Eckstein, M., \BBA\ Wang, W.~Y.
  \BBOP2022\BBCP.
\newblock \BBOQ Neuro-symbolic procedural planning with commonsense
  prompting\BBCQ.

\bibitem[\protect\BCAY{Narendra\ \BBA\ Thathachar}{Narendra\ \BBA\
  Thathachar}{1974}]{learning-automata}
Narendra, K.~S.\BBACOMMA\  \BBA\ Thathachar, M. A.~L. \BBOP1974\BBCP.
\newblock \BBOQ Learning automata - a survey\BBCQ\
\newblock {\Bem IEEE Transactions on Systems, Man, and Cybernetics}, {\Bem
  4\/}(4), 323--334.

\bibitem[\protect\BCAY{Neary, Xu, Wu,\ \BBA\ Topcu}{Neary
  et~al.}{2021}]{neary2020reward}
Neary, C., Xu, Z., Wu, B., \BBA\ Topcu, U. \BBOP2021\BBCP.
\newblock \BBOQ Reward machines for cooperative multi-agent reinforcement
  learning\BBCQ\
\newblock In {\Bem International Conference on Autonomous Agents and MultiAgent
  Systems}, AAMAS '21, \BPG\ 934–942.

\bibitem[\protect\BCAY{Neider, Gaglione, Gavran, Topcu, Wu,\ \BBA\ Xu}{Neider
  et~al.}{2021}]{Neider2021AdviceGuidedRL}
Neider, D., Gaglione, J., Gavran, I., Topcu, U., Wu, B., \BBA\ Xu, Z.
  \BBOP2021\BBCP.
\newblock \BBOQ Advice-guided reinforcement learning in a non-markovian
  environment\BBCQ\
\newblock In {\Bem {AAAI} Conference on Artificial Intelligence}, \BPGS\
  9073--9080.

\bibitem[\protect\BCAY{Petroni, Rockt{\"a}schel, Riedel, Lewis, Bakhtin, Wu,\
  \BBA\ Miller}{Petroni et~al.}{2019}]{petroni2019language}
Petroni, F., Rockt{\"a}schel, T., Riedel, S., Lewis, P., Bakhtin, A., Wu, Y.,
  \BBA\ Miller, A. \BBOP2019\BBCP.
\newblock \BBOQ Language models as knowledge bases?\BBCQ\
\newblock In {\Bem Conference on Empirical Methods in Natural Language
  Processing}, \BPGS\ 2463--2473.

\bibitem[\protect\BCAY{Pnueli}{Pnueli}{1977}]{Pnueli77LTL}
Pnueli, A. \BBOP1977\BBCP.
\newblock \BBOQ The temporal logic of programs\BBCQ\
\newblock In {\Bem Symposium on Foundations of Computer Science}, \BPGS\
  46--57.

\bibitem[\protect\BCAY{Rezaei\ \BBA\ Reformat}{Rezaei\ \BBA\
  Reformat}{2022}]{Rezaei2022UtilizingLM}
Rezaei, N.\BBACOMMA\  \BBA\ Reformat, M.~Z. \BBOP2022\BBCP.
\newblock \BBOQ Utilizing language models to expand vision-based commonsense
  knowledge graphs\BBCQ\
\newblock {\Bem Symmetry}, {\Bem 14}, 1715.

\bibitem[\protect\BCAY{Sadoun, Dubois, Ghamri{-}Doudane,\ \BBA\ Grau}{Sadoun
  et~al.}{2013}]{Sadoun2013FromNL}
Sadoun, D., Dubois, C., Ghamri{-}Doudane, Y., \BBA\ Grau, B. \BBOP2013\BBCP.
\newblock \BBOQ From natural language requirements to formal specification
  using an ontology\BBCQ\
\newblock In {\Bem International Conference on Tools with Artificial
  Intelligence}, \BPGS\ 755--760.

\bibitem[\protect\BCAY{Shah, Osinski, Ichter,\ \BBA\ Levine}{Shah
  et~al.}{2022}]{shah2022robotic}
Shah, D., Osinski, B., Ichter, B., \BBA\ Levine, S. \BBOP2022\BBCP.
\newblock \BBOQ Lm-nav: Robotic navigation with large pre-trained models of
  language, vision, and action\BBCQ\
\newblock In {\Bem Conference on Robot Learning}, \lowercase{\BVOL}\ 205 of
  {\Bem Proceedings of Machine Learning Research}, \BPGS\ 492--504.

\bibitem[\protect\BCAY{Thomason, Padmakumar, Sinapov, Walker, Jiang, Yedidsion,
  Hart, Stone,\ \BBA\ Mooney}{Thomason et~al.}{2020}]{HumanRobotDialog}
Thomason, J., Padmakumar, A., Sinapov, J., Walker, N., Jiang, Y., Yedidsion,
  H., Hart, J.~W., Stone, P., \BBA\ Mooney, R.~J. \BBOP2020\BBCP.
\newblock \BBOQ Jointly improving parsing and perception for natural language
  commands through human-robot dialog\BBCQ\
\newblock {\Bem J. Artif. Intell. Res.}, {\Bem 67}, 327--374.

\bibitem[\protect\BCAY{Vadera\ \BBA\ Meziane}{Vadera\ \BBA\
  Meziane}{1994}]{Vadera1994FromET}
Vadera, S.\BBACOMMA\  \BBA\ Meziane, F. \BBOP1994\BBCP.
\newblock \BBOQ From english to formal specifications\BBCQ\
\newblock {\Bem The Computer Journal}, {\Bem 37}, 753--763.

\bibitem[\protect\BCAY{Valkanis, Beletsioti, Nicopolitidis, Papadimitriou,\
  \BBA\ Varvarigos}{Valkanis et~al.}{2020}]{Valkanis2020ReinforcementLI}
Valkanis, A., Beletsioti, G.~A., Nicopolitidis, P., Papadimitriou, G., \BBA\
  Varvarigos, E.~A. \BBOP2020\BBCP.
\newblock \BBOQ Reinforcement learning in traffic prediction of core optical
  networks using learning automata\BBCQ\
\newblock In {\Bem International Conference on Communications, Computing,
  Cybersecurity}.

\bibitem[\protect\BCAY{Vardi\ \BBA\ Wolper}{Vardi\ \BBA\
  Wolper}{1986}]{Vardi1986AnAA}
Vardi, M.~Y.\BBACOMMA\  \BBA\ Wolper, P. \BBOP1986\BBCP.
\newblock \BBOQ An automata-theoretic approach to automatic program
  verification (preliminary report)\BBCQ\
\newblock In {\Bem Symposium on Logic in Computer Science}, \BPGS\ 332--344.

\bibitem[\protect\BCAY{Vasileiou, Yeoh, Son, Kumar, Cashmore,\ \BBA\
  Magazzeni}{Vasileiou et~al.}{2022}]{LogicExplanation}
Vasileiou, S.~L., Yeoh, W., Son, T.~C., Kumar, A., Cashmore, M., \BBA\
  Magazzeni, D. \BBOP2022\BBCP.
\newblock \BBOQ A logic-based explanation generation framework for classical
  and hybrid planning problems\BBCQ\
\newblock {\Bem J. Artif. Intell. Res.}, {\Bem 73}, 1473--1534.

\bibitem[\protect\BCAY{Vemprala, Bonatti, Bucker,\ \BBA\ Kapoor}{Vemprala
  et~al.}{2023}]{vemprala2023chatgpt}
Vemprala, S., Bonatti, R., Bucker, A., \BBA\ Kapoor, A. \BBOP2023\BBCP.
\newblock \BBOQ Chat{GPT} for robotics: Design principles and model
  abilities\BBCQ.
\newblock Published by Microsoft.

\bibitem[\protect\BCAY{West, Bhagavatula, Hessel, Hwang, Jiang, Bras, Lu,
  Welleck,\ \BBA\ Choi}{West et~al.}{2022}]{KnowledgeGraph}
West, P., Bhagavatula, C., Hessel, J., Hwang, J.~D., Jiang, L., Bras, R.~L.,
  Lu, X., Welleck, S., \BBA\ Choi, Y. \BBOP2022\BBCP.
\newblock \BBOQ Symbolic knowledge distillation: from general language models
  to commonsense models\BBCQ\
\newblock In {\Bem Conference of the North American Chapter of the Association
  for Computational Linguistics: Human Language Technologies}, \BPGS\
  4602--4625.

\bibitem[\protect\BCAY{Xiong, Du, Wang,\ \BBA\ Stoyanov}{Xiong
  et~al.}{2020}]{xiong2019pretrained}
Xiong, W., Du, J., Wang, W.~Y., \BBA\ Stoyanov, V. \BBOP2020\BBCP.
\newblock \BBOQ Pretrained encyclopedia: Weakly supervised knowledge-pretrained
  language model\BBCQ\
\newblock In {\Bem International Conference on Learning Representations}.

\bibitem[\protect\BCAY{Xu, Gavran, Ahmad, Majumdar, Neider, Topcu,\ \BBA\
  Wu}{Xu et~al.}{2020}]{Xu2020JIRP}
Xu, Z., Gavran, I., Ahmad, Y., Majumdar, R., Neider, D., Topcu, U., \BBA\ Wu,
  B. \BBOP2020\BBCP.
\newblock \BBOQ Joint inference of reward machines and policies for
  reinforcement learning\BBCQ\
\newblock In {\Bem International Conference on Automated Planning and
  Scheduling}, \BPGS\ 590--598.

\bibitem[\protect\BCAY{Zhang, Wang,\ \BBA\ Gao}{Zhang
  et~al.}{2021}]{Zhang2021LearningAM}
Zhang, Z., Wang, D., \BBA\ Gao, J. \BBOP2021\BBCP.
\newblock \BBOQ Learning automata-based multiagent reinforcement learning for
  optimization of cooperative tasks\BBCQ\
\newblock {\Bem IEEE Transactions on Neural Networks and Learning Systems},
  {\Bem 32}, 4639--4652.

\end{thebibliography}
\bibliographystyle{theapa}


\end{document}